\documentclass[twocolumn]{aastex631}
\graphicspath{{./}{GoodFigs/}}
\usepackage{float,graphicx}
\usepackage{amsmath}
\usepackage{amssymb}
\usepackage{multirow}
\usepackage{enumitem,amssymb}
\usepackage{CJK}         
\newlist{todolist}{itemize}{2}
\setlist[todolist]{label=$\square$}

\shorttitle{Long lag in Fairall 9}
\shortauthors{Yao et al.}

\begin{document}
\begin{CJK*}{UTF8}{gbsn}

\title{A Negative Long Lag from the Optical to the UV Continuum in Fairall 9}

\author[0000-0003-3024-7218]{Philippe Z. Yao}
\affil{\rm Department of Astrophysical Sciences, Princeton University, Peyton Hall, Princeton, NJ 08544, USA}

\author[0000-0002-1174-2873]{Amy Secunda}
\affil{\rm Department of Astrophysical Sciences, Princeton University, Peyton Hall, Princeton, NJ 08544, USA}

\author[0000-0002-2624-3399]{Yan-Fei Jiang (姜燕飞)}
\affil{\rm Center for Computational Astrophysics, Flatiron Institute, New York, NY 10010, USA}

\author[0000-0002-5612-3427]{Jenny E. Greene}
\affil{\rm Department of Astrophysical Sciences, Princeton University, Peyton Hall, Princeton, NJ 08544, USA}

\author[0000-0002-5814-4061]{Ashley~Villar}
\affiliation{\rm Department of Astronomy and Astrophysics, The Pennsylvania State University, 525 Davey Lab, University Park, PA 16802, USA}
\affiliation{\rm Institute for Computational \& Data Sciences, The Pennsylvania State University, University Park, PA 16802, USA}
\affiliation{\rm Institute for Gravitation and the Cosmos, The Pennsylvania State University, University Park, PA 16802, USA}


\begin{abstract}
    We report the detection of a long-timescale negative lag, where the blue bands lag the red bands, in the nearby Seyfert 1 galaxy Fairall 9. Active Galactic Nuclei (AGN) light curves show variability over a wide range of timescales. By measuring time lags between different wavelengths, the otherwise inaccessible structure and kinematics of the accretion disk can be studied. One common approach, reverberation mapping, quantifies the continuum and line lags moving outwards through the disk at the light-travel time, revealing the size and temperature profile of the disk. Inspired by numerical simulations, we expect longer lags to exist in AGN light curves that travel inward on longer timescales, tracing the accretion process itself. By analyzing AGN light curves in both temporal and frequency space, we report the detection of long-timescale lags ($\sim -70$ days) in Fairall 9 which propagate in the opposite direction to the reverberation lag. The short continuum lag ($<10$~days) is also detected and is consistent with reverberation lags reported in the literature. When fitting the longer lag as a function of frequency with a model motivated by the thin disk model, we find that the disk scale height likely increases outward in the disk. This detection raises the exciting prospect of mapping accretion disk structures across a wide range of AGN parameters.
\end{abstract}

\section{Introduction}

Supermassive black holes (SMBHs) are surrounded by an accretion disk that powers one of the most luminous sources in our universe. These central engines, termed active galactic nuclei (AGN), harbor precious information on accretion physics and galaxy evolution \citep{Fabian2012, Kormendy2013}. However, direct imaging of most accretion disks and surrounding structures is in general observationally inaccessible because of their relatively small sizes or large distance. Alternatively, temporal variability as a function of frequency can be used instead to map the structure of the accretion disks. 

Reverberation mapping is a technique that traces delays between light curves at different wavelengths. By mapping the wavelengths to different temperature regions in the disk, the lags between bands can be used to probe the accretion disk structure \citep{Blandford1982, Peterson2004}. Classically, this technique looked at the lags between AGN continua and broad emission lines to measure the sizes of the Broad Line Regions (BLR), but the same principle applies to looking for continuum lags as a function of frequency.  Since the path lengths to observers on Earth are different from different regions of the source, we can use the light travel time to deduce the physical sizes and temperature profiles of AGNs. For example, radiation from the X-ray corona gets reprocessed and re-emitted earlier in the inner disk than the outer parts. Since the former possesses much higher temperature than the latter, the reprocessed light varies in wavelength with the ultraviolet light curve leading the optical. We will refer to this lag as the “short lag” ($<10$~days), occurring in a light-crossing time, and this distinct signal has been observed in many active galaxies such as the Seyfert 1 galaxy Fairall 9 \citep[See][and references there in]{Cackett2021b}. However, signals beyond reprocessing can also be embedded in the light curves from AGN.

 In addition to the short lag, and the BLR lag, we also expect to see correlations between different bands on longer timescales, even propagating inwards in the disk. AGN accretion disks are believed to be turbulent due to magneto-rotational instability (MRI) \citep{BalbusHawley1991}. Turbulent fluctuations in the disk will naturally lead to stochastic variability from the locally emitted photons, which provide a natural explanation for the broadband photometric variability observed in AGNs. 
 When gas flows towards the black hole via the accretion disk, it is expected that turbulence-driven perturbations at larger radii will propagate inward with the accretion flow itself and modulate the local variability at small radii \citep{Lyubarskii1997}.  

Several characteristic timescales are expected from the accretion process \citep[e.g.][]{Stern2018}. Turbulence will cause light curve variability on the local thermal time scale for optically thick disks \citep{JiangBlaes2020,Burkeetal2021}. When perturbations propagate inward, fluctuations at different radii can correlate on the inflow time scale. Particularly, in AGN accretion disks, the UV/Optical emission region is expected to be strongly perturbed by convection due to enhanced opacity \citep{JiangBlaes2020} in addition to the normal MRI turbulence, which can enhance the fluctuation propagation signals. Since the effective temperature at large radii should be lower than the effective temperature of the inner region of the disk, this kind of fluctuation propagation can cause long wavelength variability to lead the variability at shorter wavelengths, which will be in the opposite direction as traditional reverberation signals. The two kinds of correlation will happen on quite different timescales but both of them (and potentially the BLR lag, which is not treated here) will exist in the light curve. Despite continuum lags, BLR lags also exist in AGN light curves, but are not the focus of this paper. Variability studies of X-ray binary light curves already find evidence that both lags can exist in the very inner region of the accretion disks \citep{uttley2014}. Inward fluctuation propagation has also been proposed to explain the variability properties of many accretion systems \citep[e.g.][]{Arevalo2006, Arevalo2008a,Arevalo2008b, HoggReynolds2016}. Recent variability analysis of AGN has identified systems with two components with different timescales \citep[e.g.][]{F92020, Vincentelli2021}, but the exact long-timescale lag remains uncertain. Inflow time scales in AGN accretion disks will depend on the detailed structure of the disk, and our a priori expectation is that these can be $\sim 100-1000$ days, but more empirical constraints are needed. Detecting such lags in AGN light curves will be very useful to probe internal structures of AGN accretion disks, and is a main goal of this work. 

In this paper, we explore both cross-correlation \citep{Gaskell1987} and frequency resolved techniques to extract AGN time lags with specific interest in extracting a long lag in Fairall 9. In temporal space, we use the JAVELIN code (\citet{Zu2010}, formerly known as SPEAR), which uses a damped random walk (DRW) model \citep{Timmer1995} to describe the AGN variability, and interpolates the light curves to increase sensitivity to the short-timescale variability. In frequency space, the Fourier method can provide excellent information on different timescales and periodicity. Fourier methods have seen many successes in X-ray astronomy, such as characterizing X-ray binaries. Beyond a single lag, frequency-resolved analysis yields lag as a function of frequency and can find the characteristic timescales of different processes. However, this technique can be challenging when irregular sampling is present. Both high and low frequency lags in X-ray reverberation of AGN have been detected through simple fast Fourier transform with continuously sampled data \citep[e.g.][]{Papadakis2001, McHardy2007, Arevalo2008a, Fabian2009, Zoghbi2010, DeMarco2013, Cackett2013, uttley2014}. Observations of AGN in the UV/optical require a maximum-likelihood approach in frequency space because of their non-uniform cadence. Hence, we use the method developed by \citet{Miller2010}, which \citet{Zoghbi2013} and \citet{Cackett2021} later adapted to successfully recover lags from AGN light curves. We investigate the relative strengths of each method for extracting a range of lag times in a single spectrum. This analysis will be carried out both on simulated AGN light curves and on multi-wavelength observations of the local Seyfert galaxy Fairall 9, which hosts a $2.55 \times 10^8$~M$_{\rm \sun}$ SMBH \citep{Peterson2004}.

Our discussion begins with presenting one of the best existing data sets for our studies from \citet{F92020} in Section \ref{sec:data}. Then we move on to a introduction to both the cross-correlation and Fourier methods in Section \ref{sec:peda}. Next, in Section \ref{sec:F9}, we report our recoveries of both the long and short lags from the Fairall 9 data set, which informs how future long-term monitoring of AGN will improve our understanding of both signals. We then discuss how the wavelength dependence of the long lag informs the geometry of the inner accretion disk in Section \ref{sec:discussion}. Finally, we provide a brief summary in Section \ref{sec:summary}. Details regarding the applicability of both methods in different scenarios are presented in Appendix \ref{app:sim}, showing lag recovery from simulated light curves following a DRW model.

\pagebreak

\section{High Cadence Monitoring of AGN} \label{sec:data}
With continuum reverberation mapping, significant progress has been made in understanding accretion physics, with pioneering observational projects such as the Space Telescope and Optical Reverberation Mapping Project (STORM) \citep[e.g.][]{derosa2015, Edelson2015, Fausnaugh2016, Starkey2017}), and other intensive disk-reverberation mapping campaigns \citep[e.g.][]{Edelson2017,Edelson2019,Cackett2018,McHardy2018,F92020}. Low-cadence sampling ($\sim$ 2-7 days) and short wavelength coverage are often the causes for failing to detect inter-band lags in analysis of AGN light curves \citep{F92020}. In a handful of AGN, high sampling, wide wavelength coverage, and long baseline are all available, allowing simultaneous studies of both short-timescale ($<10$~days) and long-timescale ($\sim$ 100 days) lags.

For our analysis, we focus on the intensive disk-reverberation mapping campaign of Fairall 9 carried out by \citet{F92020} using \emph{Swift} and the Las Cumbres Observatory (LCO). Half of the data are already released to the public, hence we use light curves covering half of the 1.7-year duration between MJD 58250-58550. This timespan is still much longer than the timescale at which the short lag occurs. These data are also characterized by extended wavelength coverage from 1928 $\rm \AA$ (\emph{Swift} $UVW2$) to 8700 $\rm \AA$ (LCO $z_s$). In the UV/optical regime, with $UVW2$, $UVM2$, $UVW1$, $U$, $B$, $V$ broad-band filters from \emph{Swift} and $uBVg'r'i'z_s$ filters from LCO, the wavelength dependence of both lags can be more clearly identified and associated with different regions of the accretion disk. Additionally, light curves from this campaign have a one-day cadence from \emph{Swift} and a sub-day cadence from LCO, allowing more accurate derivation of the shorter lag. 

In the intensive disk-reverberation mapping campaign of Fairall 9, \citet{F92020} were able to derive robust short lags using a cross-correlation method and provided a wavelength dependence consistent with $\tau \propto \lambda^{4/3}$ from classic predictions. For Fairall 9, with a Eddington ratio $L/L_{Edd} = 0.02$, the increase in lag from $UVW2$ to \emph{$z_s$} band shows promising compatibility with accretion disk theory. Additionally, the authors identified a low-frequency and slowly-varying component in Fairall 9 evidenced by a parabolic shape over $\sim$ 100 days. By measuring the shift in the minima of each band's parabolic fit, they derived a $\lesssim 10$ day lag between the blue and red bands, which they conclude corresponds to the propagation time convolved with a weighted radial emission profile in different regions. \citet{F92020} associated this slower component with disk fluctuations, and left further analysis for future work when longer and higher-cadence monitoring of Fairall 9 becomes available.  The reported lag ($\sim 10$ days) for this slowly varying component is even smaller than the local thermal time scales in this region of the disk, which is one important motivation for us to reanalyze the light curves.

Frequency-resolved analysis could identify different processes taking place on different timescales \citep{uttley2014}, and has not been carried out on this data set. With unevenly sampled light curves, \citet{Zoghbi2013} builds upon the Fourier method presented first in \citet{Miller2010}, to successfully recover lags from NGC 4151 and MCG-5-23-16 that are consistent with standard cross-correlation methods. As a result, we are motivated to explore the timescale at which the slowly varying component seen in the intensive disk-reverberation mapping campaign of Fairall 9 occurs by assessing the ability of both the cross-correlation and Fourier methods in simultaneously constraining both varying components from the driving and the reprocessed light curves. 

After demonstrating both methods in the following sections, we will report the first direct confirmation of the existence of a negative long-timescale lag ($\sim -70$ days) in Fairall 9 in Section \ref{sec:F9} obtained through cross correlating an interpolated light curve with JAVELIN.

\section{Finding Lags}\label{sec:peda}
    Continuum reverberation mapping aims at measuring the structure of AGN accretion disks, and hence determining properties of the black hole. Doing so relies on time-series in multiple bands where time lags between light curves across a wavelength range can be measured. In recent years, both observational and theoretical work \citep[e.g.][]{F92020,Neustadt2022} has revealed the potential existence of the long lag inherent to accretion fluctuations in AGN UV/Optical light curves, especially in Fairall 9. As a result, in this section we build upon the methods of cross-correlation and Fourier to simultaneously constrain the short and long lags. 
    \subsection{Cross-correlation lags} \label{sec:CC} 
    With an ideal uniformly-sampled time-series, cross-correlation analysis can be easily performed by measuring the cross-correlation coefficient at each delay-time $\tau$ following:
    \begin{equation}\label{eq1}
    \rm CC_{\rm coeff}(\tau) = \frac{\sum d(t)r(t+\tau)}{\sqrt{\sum d(t)^2 r(t+\tau)^2}},
    \end{equation}
    where d(t) and r(t) represent light curves from different bands used for cross-correlation; in this equation, r(t) is assumed to be delayed by time $\tau$ compared to d(t). The highest coefficient indicates the most probable time lag. Both positive and negative time lags can be measured in this manner by selecting a range of $\tau$ with both signs. The accuracy of this method is largely dependent on the discrete sample cadence. However, without the premise of uniform cadence, simple cross-correlation can no longer easily recover accurate lags, and interpolating to a uniform light curve becomes the natural solution. 
    
    JAVELIN \footnote{As presented in \citet{JAVELIN}, code available at: \url{https://github.com/nye17/javelin}}, previously known as SPEAR, is an algorithm used to look for time lags in spectroscopic \citep{Zu2011} or photometric \citep{Zu2016} light curves. To perform the latter analysis, JAVELIN models quasar light curves DRW stochastic process. This has been shown to be an effective representation of quasar optical variability on timescales from days to years \citep[e.g.][]{Kelly2009, Kozlowski2010, Zu2013}. By using Markov Chain Monte Carlo (MCMC) to estimate model parameters, JAVELIN averages over all possible continuous light curves consistent with the input data and the DRW model while simultaneously fitting a transfer function to maximize the likelihood function. To calculate the most probable time lags, JAVELIN takes into account multiple sources of error including those associated with interpolation. When applied to our simulated light curves and real data, JAVELIN takes the epoch, flux, and uncertainty as inputs, and returns the parameters for the interpolated DRW light curve, as well as the probability distribution for time lags. By selecting the range of lags we are looking for, both positive and negative, JAVELIN can constrain both lags simultaneously.
    
    \begin{figure*}[ht]
        \centering
        \includegraphics[width=\textwidth]{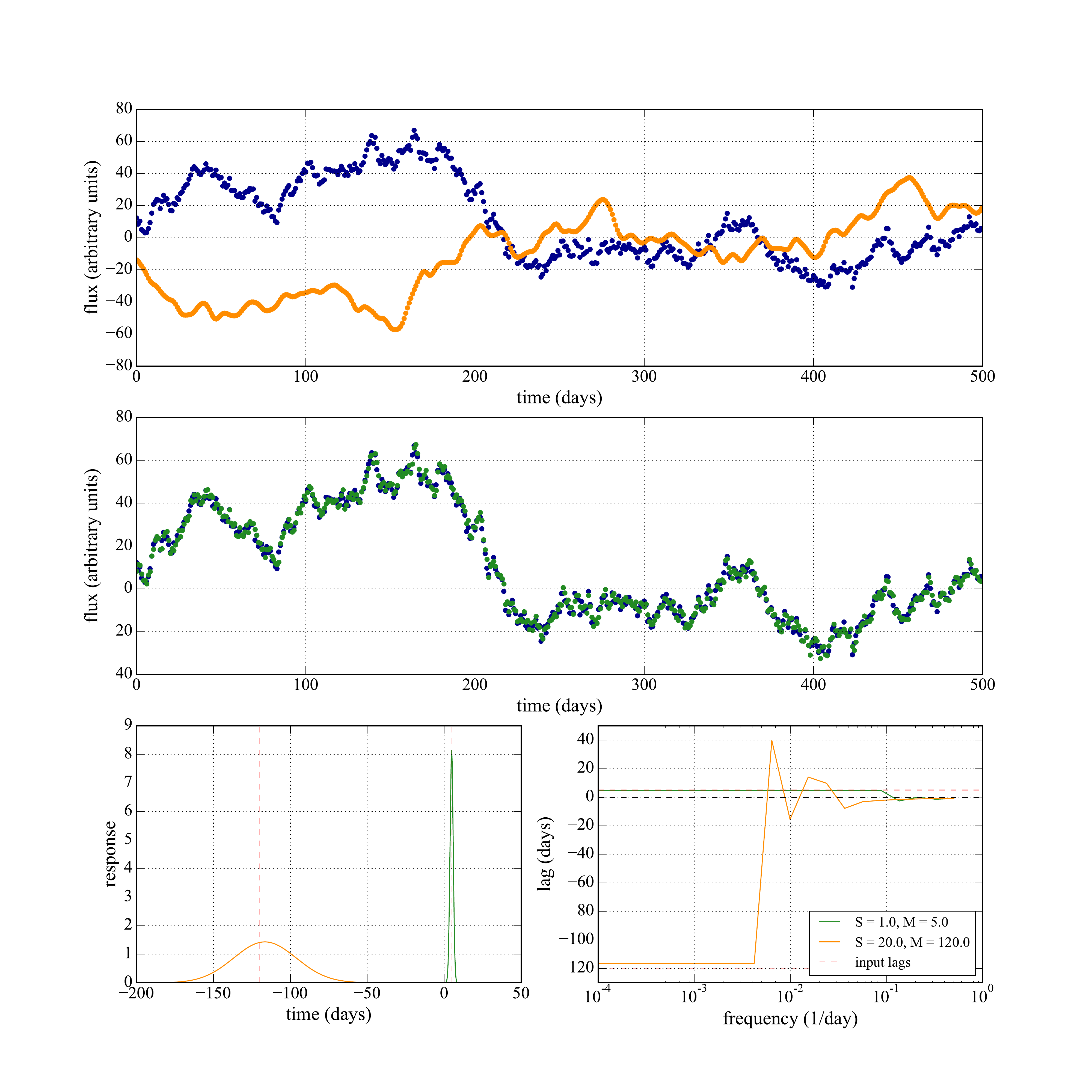}
        \caption{Top panel: simulated DRW light curve (blue) with uniform cadence, and reprocessed light curve with a long lag (orange). Middle panel: simulated DRW light curve (blue) with uniform cadence, and reprocessed light curve with a short lag (green). Each light curve has a length of 500 days and is sampled once per day. Lower left: response functions that contains either a long lag (orange) or short lag (green). The ratio of the amplitudes for each lag is set to be 1:5 for test purposes. Lower right: recovered long (orange) and short (green) lags with frequency-resolved lag analysis. In this ideal case, both lags can be easily recovered and identified by the flattening of lags at lower frequency.}
        \label{fig:freq}
    \end{figure*}
    
    \subsection{Frequency-resolved lags} \label{sec:freq}
    Fourier transforms naturally separate variability at different frequencies. In the regime of X-ray binaries, frequency-resolved analysis distinguishes different processes taking place on different timescales \citep{uttley2014}. Due to the short time scale of physical processes in this context, very uniform cadence can be achieved at X-ray wavelengths where the disk is most luminous. However, when applied to UV/optical light curves from AGNs, the relevant physical timescales are much longer, and the disk emission is spread out from UV through optical. As a result, uniform cadence is never fully realized, and more delicate treatment is necessary for an accurate lag determination.

        \subsubsection{Uniform cadence time-series}

        Reprocessing of the driving light curve $d(t)$, as shown in the top panel of Figure \ref{fig:freq}, can be represented as a response function $\psi(t)$ where the reprocessed light curve in a linearized reveberation mapping model is
        \begin{equation}\label{eq2}
        \rm r(t) = \int_{-\infty}^{\infty}\psi(\tau)d(t-\tau)d\tau.
        \end{equation}        
        In the Fourier domain, this integral becomes a multiplication in the form of 
        \begin{equation}\label{eq3}
        \rm R(\nu) = \Psi(\nu)D(\nu),
        \end{equation}   
        where upper-case variables represent the Fourier transform of their lower-case counterparts at each frequency $\nu$. The power-spectral density (PSD) of the driving light curve is simply equal to $\rm |D(\nu)|^2 = D^*(\nu)D(\nu)$, and the PSD for the reprocessed light curve follows in the same manner via $\rm |R(\nu)|^2 = |\Psi(\nu)|^2|D(\nu)|^2$.  
        Then, the cross-spectrum of the two light curves can be measured as the product of the complex conjugate of one Fourier-transformed light curve and the Fourier transform of the other:
        \begin{equation}\label{eq4}
        \rm C(\nu) = D^*(\nu)R(\nu) = |D(\nu)|^2\Psi(\nu). 
        \end{equation}          
        Following \citet{uttley2014} further, one can calculate the phase lag $\phi$ as the phase of the complex cross spectrum, and derive the time lag $\tau$ through
        \begin{equation}\label{eq5}
        \tau = \phi/2\pi\nu.
        \end{equation}        
        As seen in Figure \ref{fig:freq}, the resulting distribution of $\tau$ in frequency space is largely dependent on the shape of the response function. Unlike the case for the single short lag, even when the light curve is reprocessed just with a single long lag, it can be difficult to discern by eye. But with the Fourier method, we can recover the lags in both cases indicated by the constant lag in the flat part of the lag distribution at lower frequencies. At higher frequencies, we see expected phase-wrapping as a form of aliasing. Phase wrapping depends on the centroid of the response function ($\tau_{\rm cent}$) at the frequency $\nu = 1/2\tau_{\rm cent}$. Because the waveform could be shifted by half a wave, the lag distribution in frequency space wraps around from $\pi$ to $-\pi$, from positive to negative \citep{uttley2014}.

        \subsubsection{Non-uniform cadence time-series} 
        With UV/optical AGN light curves, the uniform sampling available from X-ray satellites is no longer feasible, at least at present. As a result, lag determination in frequency space cannot rely on simple fast Fourier transforms as presented above. Thus, a Fourier method with a maximum-likelihood approach, first presented in \citet{Miller2010}, is more suitable. In this case, we model the power spectra and response functions as piecewise functions of Fourier frequency to construct a likelihood function, which is then used to estimate model parameters when maximized. For all of our analysis of simulated and real data, we follow the detailed outline of the method presented by \citet{Zoghbi2013}.
        
        We begin by creating N bins in frequency space with ranges and centroids depending on the simulated light curve length. The bins range from $\rm f_{min} = 1/T$ to $\rm f_{max} = 1/(2dt)$, where T is the longest separation between points in the light curves and dt is the shortest. Two additional bins are added at the start and end of the other bins to reduce bias in the first and last bins. We then create a model for the PSD and use a quadratic approximation to the likelihood to maximize the likelihood function and find best fit values of the PSD. Our uncertainties come from the Hessian matrix of the likelihood function, which places a lower limit. Future work can estimate the full uncertainty using the emcee package if needed. This step is repeated for each light curve and both PSDs are then used as inputs for modeling the cross spectrum to calculate the phase delay. 
        
        Here, the autocorrelation function can be calculated via:
        \begin{equation}\label{eq6}
        \rm  \mathcal{A}(\tau) = \int_{-\infty}^{\infty}|D(\nu)|^2cos(2\pi\nu\tau) d\nu,
        \end{equation}   
        where $\rm |D(\nu)|^2 = \sum_i D_i$ is the PSD as a piecewise sum over a range of frequency bins. Then, the time lag can be estimated with the cross-covariance, whose relationship with the cross spectrum and phase lag are embedded in the cross-correlation function:
        \begin{equation}\label{eq7}
            \begin{split}
                \rm \chi(\tau) &= \int_{-\infty}^{\infty}  {\rm C}(\nu){\rm cos}(2\pi \nu \tau-\phi(\nu)) d\nu \\
                            &= \int_{-\infty}^{\infty}|\rm D(\nu)|^2|\Psi(\nu)|\rm cos(2\pi\nu\tau-\phi(\nu)) d\nu,
            \end{split}
        \end{equation}          
        where $\Psi(\nu) = \sum_i \Psi_i$ and $\phi(\nu) = \sum_i \phi_i$ are the Fourier transform of the response function and its phase. Thus, the algorithm can construct and maximize a likelihood function to optimize for $\Psi_i$ and $\phi_i$ at each frequency bin. The resultant $\phi_i$ distribution can be further divided by $2\pi\nu_i$ to calculate the time lag in frequency space, producing a lag distribution similar to the uniform cadence case with frequency-resolved analysis. The intervals of log(likelihood) informs the error: the 68\% uncertainty associated with each parameter can be estimated as the value that changes $\rm -2log(\mathcal{L}/\mathcal{L}_{max})$ by 1 while allowing other parameters to vary \citep{Miller2010}. In our case, with a small set of parameters to optimize, this is computationally inexpensive. However, with $\gtrsim 20$ parameters, one can use MCMC to map the whole probability space. 

    \subsubsection{Simulated light curves}
        We test both the Fourier (frequency space) method and the cross-correlation method (temporal space) with simulated DRW light curves, using JAVELIN for the latter. Our mock light curves for testing are inspired by the Fairall 9 data set in signal-to-noise ratio and cadence, but then we explore the impact of changes to the length of the campaign, as well as recovery success as a function of the shape of the response function. More details related to our simulated light curves and performance of both methods on extracting time lags with both uniform and non-uniform time-series are presented in Appendix \ref{app:sim}.
        
        In summary, both methods show significant promise in constraining long- and short-timescale lags simultaneously in non-uniform cadence light curves. With a cadence as high as the Fairall 9 data set ($\sim$ day to subday), both methods can easily recover the short lag with low uncertainties. For the long lag, the Fourier method requires the time baseline of observations to be $\gtrsim 6$ times the long lag ($\sim$ 2 years in the case of a -120 day lag ) to probe into the relevant time scale in frequency space; JAVELIN can recover the long lag given the length of current Fairall 9 campaign ($\sim 300$ days) but lacks information regarding the physical timescale. Since the actual shape of the response function around AGN is generally unknown, our test cases adopt several amplitude ratios between the responses of the short and long lags. As expected, when the response function for the long lag is stronger, the lags are more easily recovered, and vice versa.   
        
        Both JAVELIN long lags and Fourier short lags show no evidence of contamination from the other lag. However, JAVELIN short lags and Fourier long lags can be biased by the other lag to different degrees depending on the relative amplitudes of the response functions. This could provide a remedy to a well-known puzzle with continuum reverberation mapping that accretion disk sizes derived from cross-correlation lag spectrum are often $\sim 2-3$ times bigger than expected by the standard accretion theory in  \citet{ShakuraSunyaev1973} \citep[e.g.][]{Edelson2015,Edelson2017,Edelson2019,Fausnaugh2016,Cackett2018, Pozo2019}. One possible explanation is that the value of the short lag can be biased by other embedded signals in AGN light curves. Figure \ref{fig:res_amp_JAV} in Appendix \ref{app:sim} shows that lags recovered with cross-correlation can be notably longer than the input lag when the signal of the long lag is more significant, potentially resulting in inaccurate measurements of disk sizes. In this case, better modelling is needed to accurately deduce accretion disk sizes, as shown in \citet{Kammoun2021}. On the other hand, Figure \ref{fig:res_amp} shows that short lags deduced with the Fourier method are all consistent with the input lag at 5 days, regardless of the relative amplitude ratio of the response functions. This corroborates our previous conclusion that the Fourier method tends to recover more unbiased and precise measurements of the short lag, which provides a hopeful outlook in accurately measuring accretion disk sizes across a large AGN population.

\begin{table*}[ht]
    \centering
        \begin{tabular}{cccccc}
            \hline
            Filter & $\lambda_{\rm eff}$($\rm \AA$) & Fourier (short) & JAVELIN (short) & JAVELIN (long) \\
            \hline
            \textit{Swift}\\
            $M2$    & 2246 & $0.34 \pm 1.33$ & $0.23 \pm 0.39$ & $-29.6 \pm 21.9$\\
            $W1$    & 2600 & $-1.20 \pm 1.28$ & $1.94 \pm 1.16$ & $-43.5 \pm 15.9$\\
            $U$     & 3465 & $-0.18 \pm 1.20$ & $3.56 \pm 1.45$ & $-65.6 \pm 3.4$\\
            $B$     & 4392 & $1.86 \pm 0.69$ & $6.02 \pm 3.73$ & $-58.5 \pm 5.4$\\
            $V$     & 5468 & $0.39 \pm 1.32$ & $6.94 \pm 2.25$ & $-71.4 \pm 8.5$\\
            \hline
            LCO \\
            $u'$    & 3580 & $1.73 \pm 0.40$ & $4.41 \pm 0.61$ & $-58.6 \pm 9.8$\\
            $B$     & 4392 & $1.26 \pm 1.20$ & $5.62 \pm 0.63$ & $-65.9 \pm 4.6$\\
            $g'$    & 4770 & $1.04 \pm 0.72$ & $3.08 \pm 0.96$ & $-66.9 \pm 7.4$\\
            $V$     & 5468 & $1.47 \pm 4.71$ & $4.62 \pm 0.82$ & $-69.1 \pm 9.9$\\
            $r'$    & 6215 & $2.22 \pm 0.71$ & $4.44 \pm 1.70$ & $-59.2 \pm 6.7$\\
            $i'$    & 7545 & $2.61 \pm 0.66$ & $8.77 \pm 2.16$ & $-77.3 \pm 9.8$\\
            $z_s$   & 8700 & $3.30 \pm 1.06$ & $6.93 \pm 1.53$ & $-68.0 \pm 6.2$\\
            \hline
        \end{tabular}
        \caption{Table of lag values as plotted in Figure \ref{fig:F9lags}. Note that JAVELIN short lags come from detrended light curves. All lags are measured in units of days, relative to the reference $UVW2$ band (1928 $\rm \AA$).}
        \label{tab:lags}
\end{table*}

\section{Lags in Fairall 9 Time-series}\label{sec:F9}
    
    Building on the work of \citet{F92020}, we revisit the Fairall 9 UV/Optical light curves. Specifically, we hope to investigate whether the low-frequency component they uncovered may correspond with the long lag we expect. 
    To make comparison easier, we show the lag calculations in each band relative to the $UVW2$ band (far-UV from \emph{Switft}) as in \citet{F92020}. 
    
\begin{figure}[H]
        \centering
        \includegraphics[width=\columnwidth]{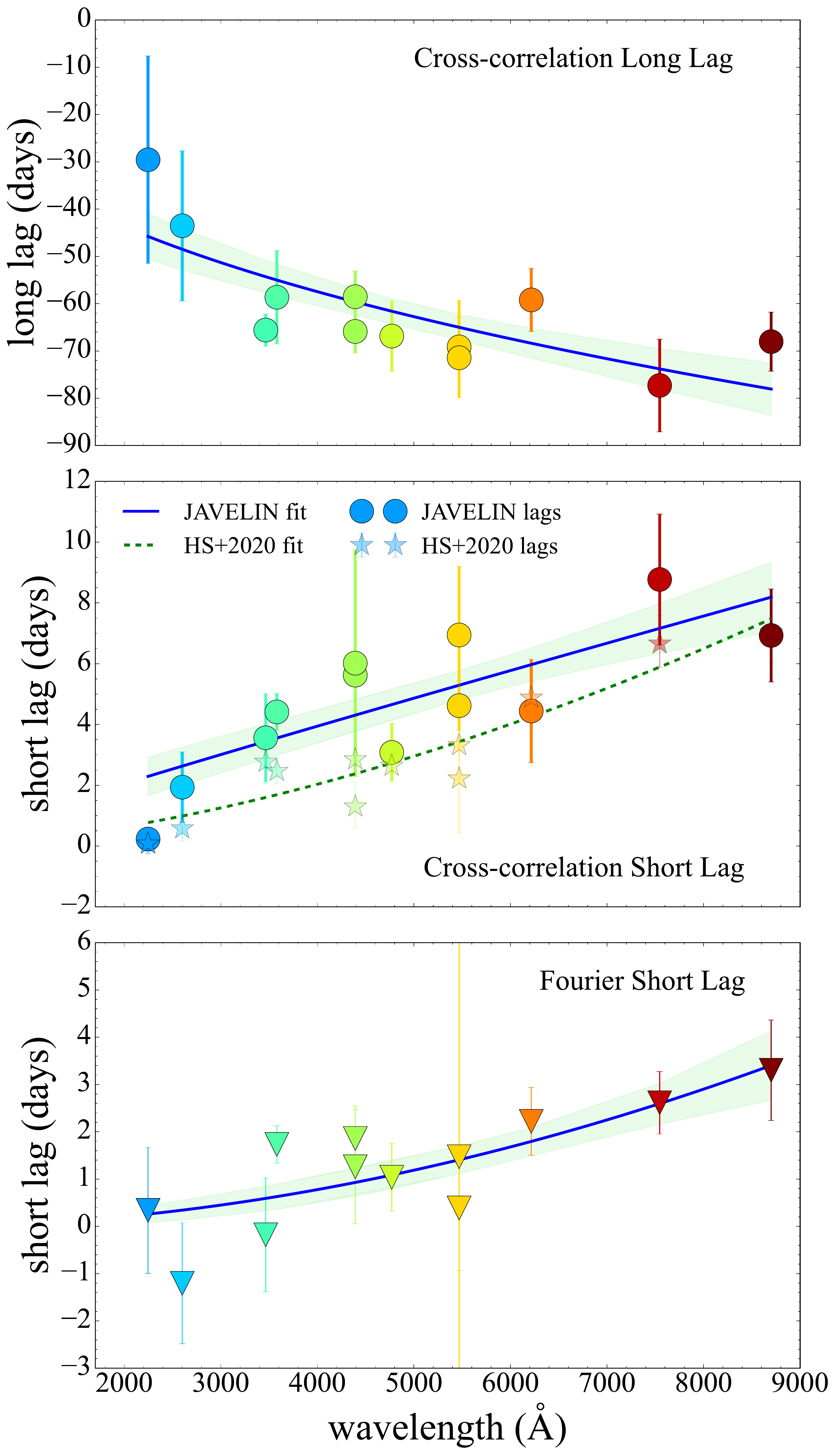}
        \caption{JAVELIN long lags (top) and detrended short lags (mid), and Fourier method short lags (bottom) recovered from the Fairall 9 data as a function of wavelength. Light curves are obtained from \citet{F92020}. The $UVW2$ band is used as a reference band to perform cross-correlation against by JAVELIN. Long lags $\sim -70$ days are detected in all bands with a sign opposite to the short lags. All lags are increasing in magnitude with wavelength. The simple power law fits with 1$\sigma$ uncertainty are also shown as the blue solid line and the green shaded region. Note that in JAVELIN the long lag in the $M2$ band (2246 $\rm \AA$, \textit{Swift}) is measured separately (for the cross-correlation functions see Figure \ref{fig:F9_jave}, and Figure \ref{fig:M2_neg} for the long lag in M2). For easier comparison, cross-correlation short lags from \citet{F92020} (labeled as HS+2020) are overplotted as stars in the middle panel with the same power law fit in dashed line, where circular points are JAVELIN short lags from detrended light curves.}
        \label{fig:F9lags}
    \end{figure}

    For analysis of short lags, detrending AGN light curves can remove variability on long timescales and hence produce more accurate measurements of the short timescale lags \citep{McHardy2014,McHardy2018,Pahari2020}. Therefore, we perform quadratic detrending to analyzing the Fairall 9 data as in \citet{F92020} when measuring the short lag with JAVELIN. Our results also show short lags from undetrended light curves are about 2 times shorter than from detrended light curves. As seen in the middle panel of Figure \ref{fig:F9lags}, the short lag is increasing in magnitude with wavelength as predicted. In the bottom panel, we also show the detected short lags from the Fourier method. These lags come from undetrended light curves, since the Fourier approach should naturally separate variability on different timescales. When compared to the cross-correlation results, we observe a recovery of the short-timescale lag that is better-fit with a power law and has a generally smaller error compared to the cross-correlation method. The success in this regime by the Fourier method can be attributed to the fact that the short lag occurs at a timescale significantly shorter than the full length of the available Fairall 9 data, and that interpolation is a noticeable source of error in JAVELIN. This leads to a conclusion that high-cadence (subday to $\sim$ 1-day) monitoring of AGN will significantly benefit the recovery of the short lag, also highlighting the strength of the Fourier method when the sampling and length of light curve is adequate. 
    
    For the long lag, when using the Fourier method, we can only probe frequencies as long as half of the time span. Hence, similar to our test case when we create a simulated Fairall 9 data set that extends for only $\sim 300$ days with a 120-day lag injected (see Appendix \ref{app:sim}), our result is largely limited by the half-length of the Fairall 9 light curves and we expect improvements when the full 1.7-year Fairall 9 light curves become publicly available. This is further bolstered by the fact that our simulations in Appendix \ref{app:sim} show that light curves with $\gtrsim$ triple the time span of the $\sim 300$-day light curves allow for more accurate long lags from the Fourier method. Hence, long-baseline observations are especially important for detecting long lags in Fourier space. However, when using JAVELIN on real data, we robustly detect long lags in almost all wavelengths except the $M2$ band immediately adjacent to the reference $UVW2$ band, as seen in the top panel of Figure \ref{fig:F9lags}. For the $M2$ band, we recover a long lag with relatively large uncertainties when limiting JAVELIN to search for only negative lags. Since the wavelength difference between $M2$ and $UVW2$ are small, the long lag is much shorter, and it is possible that the short lag overwhelms the long lag. A similar $\sim -30$-day lag is also seen in the Fourier approach, as in the first panel of Figure \ref{fig:Fourier_lags}. But we did not explore this in detail, primarily due to the limited length of the light curves available for the Fourier method. Both JAVELIN short and long lags and Fourier short lags are reported in Table \ref{tab:lags}.The exact cross-correlation function for each inter-band lag analysis with JAVELIN is available in Figure \ref{fig:F9_jave}. The cross-correlation functions for the detrended light curves analyzed with JAVELIN are available in Figure \ref{fig:F9_de_jave}. The frequency-space lag distributions are available in Figure \ref{fig:Fourier_lags}.
    
    As expected, the long lags are all negative, which indicates an opposite propagation direction to the short lag. The lag magnitudes range from $\sim -30$ to $\sim -80$ days, increasing with wavelength until $\rm 4392\, \AA$ and then gradually flattening. In Section \ref{sec:discussion}, we will discuss the physical picture associated with this dramatic increase in the lag within regions closer to the black hole. This recovered long lag is a better match for theoretical predictions of the slowly varying component, which separates itself with both lag direction and magnitude from the short lag occurring at the light-crossing time. Figure \ref{fig:res_amp_JAV} in Appendix \ref{app:sim} shows that regardless of the relative amplitudes of the long and short lag, the long lag can be reliably recovered.
    Hence, here we do not expect the value of the long lag we recover to be significantly biased by the short lag. 
    
    To confirm that this long lag is not spurious, we conduct the same analysis with JAVELIN using simulated inputs of a single short lag sampled at the same epochs as the real data. In this case, no long lag is detected in JAVELIN. Additionally, when the real data are shuffled, no lag is detected in the cross-correlation function. Further tests with DRW models without reprocessing are discussed in Appendix \ref{app:sim}, along with an example of the original and detrended Fairall 9 light curves in Figure \ref{fig:lcs}.
    
    To compare, \citet{F92020} reported the long lags with a $\sim 100$-day timescale of variability but with a negative time lag of only $\lesssim 10$ days in magnitude between the $UVW2$ and $z_s$ bands. This analysis is carried out by performing a quadratic fit to the light curves, and then using the time shift among minima of corresponding parabolas at each band as an indication of the time lag.  The fitting is based on this single peak, which introduces significant error. For Fairall 9, our reported long lag values at $\sim -70$ days differ by more than an order of magnitude from the previous method.  The lag value we find is also more consistent with the expected inflow timescale if this lag is produced by inward propagation of disk fluctuations.
    Further discussion regarding the power law fitting and how this result informs the characteristics of the accretion disk are included in Section \ref{sec:discussion}.

\section{Wavelength dependence of the long lag}\label{sec:discussion}

\begin{figure*}[ht]
        \centering
        \includegraphics[width=0.77\textwidth]{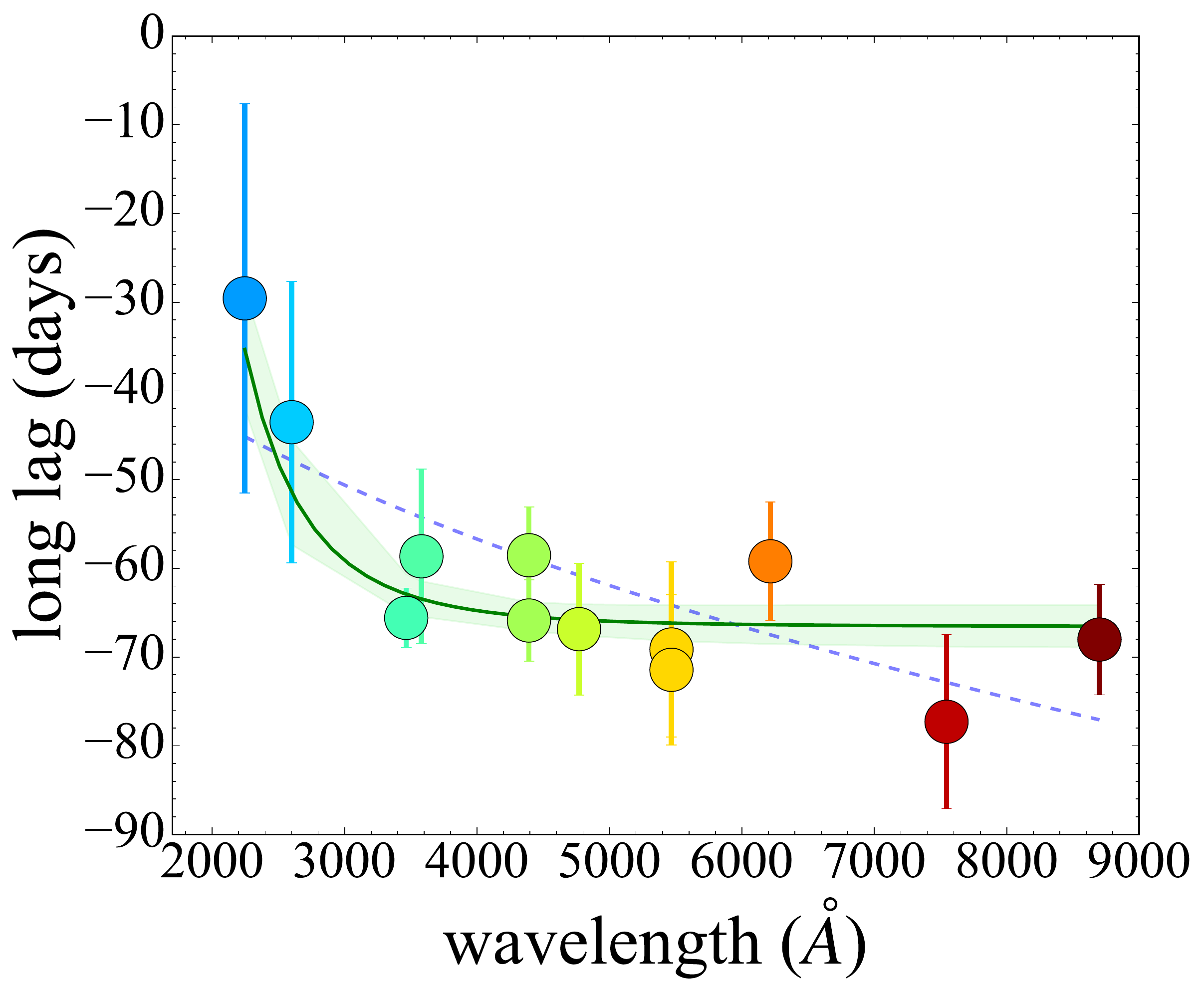}
        \caption{Long lags recovered from Fairall 9 as in Figure \ref{fig:F9lags} but fitted by forcing the power law to go to zero at the reference wavelength. The green shaded region reflects the 1$\sigma$ uncertainty of this fit. The blue dotted line shows the simple power law fit that resembles the fit in Figure \ref{fig:F9lags} in the form of $\tau = \tau_0(\lambda /\lambda_0)^\gamma$ }
           \label{fig:F9_LL_PL}
\end{figure*}

\begin{table*}[ht]
    \centering
        \begin{tabular}{cccccc}
            \hline
            Lag Type & Model Description & $\tau_0$ & $\gamma$ & $\chi^2$ & Reduced $\chi^2$ \\
            \hline
             \multirow{2}{*}{Long lag} & simple power law &  $-43.1 \pm 4.9$ &  $0.39 \pm 0.11$ &  751.20 & 75.12\\
             & simple power law (forced to zero) &  $66.8 \pm 2.3$ & $-4.90 \pm 1.41$ &  5.73 & 0.57\\
            \hline
             \multirow{3}{*}{Short lag} & detrended JAVELIN &  1.99 $\pm$ 0.59 &  0.94 $\pm$ 0.26 &  22.3 & 2.23\\
             & Fourier &  $0.19 \pm 0.14$ &  $1.90 \pm 0.57$ & 6.42 & 0.64\\
             & Fourier (forced to zero) &  $0.90 \pm 0.64$ &  $0.88 \pm 0.47$ & 8.01 & 0.80\\
             \hline
        \end{tabular}
        \caption{Best-fitting parameters of simple power law models for the long and short lags in Fairall 9. For the long lag, only the JAVELIN results were analyzed, whereas for short lags, both JAVELIN and Fourier results were fitted.}
        \label{tab:results}
\end{table*}

At first glance, the steep wavelength dependence of the long lag seems suspicious. Here we will argue that in principle there is a physical explanation for it, namely radial variation in the disk thickness. To better understand the implications of our lag distributions as a function of wavelength, first we use the simple functional form of a power law to fit the lags from both methods:
 \begin{equation}
    \tau = \tau_0\left(\frac{\lambda}{\lambda_0}\right)^\gamma, 
 \end{equation}
where $\tau$ is the recovered lag, $\tau_0$ is the normalization constant, $\lambda_0$ is the reference wavelength band at 1928$\rm \AA$, and $\gamma$ is the power law slope. For the detrended short lag, we obtain $\tau = (1.99 \pm 0.59)\times(\lambda/\lambda_0)^{0.94 \pm 0.26}$ from JAVELIN and $\tau = (0.19 \pm 0.14)\times(\lambda/\lambda_0)^{1.90 \pm 0.57}$ from the Fourier method, where wavelengths are measured in Angstrom. Alternatively, we can use a fiducial wavelength $\rm \lambda_c \equiv 5468 \AA$ at the center of our wavelength range, and force the lag at reference wavelength ($1928 \AA$) to be zero \citep[e.g.][]{Pal2017} with
 \begin{equation}
    \tau = \tau_0\left[\left(\frac{\lambda}{\lambda_c}\right)^{\gamma}-\left(\frac{\lambda_0}{\lambda_c}\right)^{\gamma}\right]. 
 \end{equation}
Having $\rm \lambda_c$ gives more flexibility to the fit. The short lag recovered from the Fourier method can be fitted with $\tau = (0.90 \pm 0.64)\times[(\lambda/{\lambda_c})^{0.88 \pm 0.47}-(\lambda_0/\lambda_c)^{0.88 \pm 0.47}]$. Both measurements of the power law slope with the Fourier short lags are consistent with the standard accretion disk model scaling of $\tau \propto \lambda^{4/3}$, while it lies slightly outside of 1$\sigma$ in the JAVELIN result.

For the long lag from JAVELIN, we recover a relationship of $\tau = (-43.1 \pm 4.9)\times(\lambda/\lambda_0)^{0.39 \pm 0.11}$ with the simple power law fit. With a reduced $\chi^2 = 75.12$, this simple power law is not a good fit to the long lag. On the other hand, we can force the fit to go to zero at the reference wavelength with:
 \begin{equation}
    \tau = \tau_0\left[\left(\frac{\lambda}{\lambda_0}\right)^{\gamma}-1\right],
 \end{equation}
with $\lambda_0 = 1928 \rm \AA$. In this case, we recover a relationship of $\tau = (66.8 \pm 2.3)\times[(\lambda/\lambda_0)^{-4.90 \pm 1.41}-1]$ with a reduced $\chi^2 = 0.57$ as illustrated in Figure \ref{fig:F9_LL_PL}. 

Given our knowledge of accretion disk theory, we can derive a more sophisticated functional form for the long lag to interpret the meaning of the power law slope. Long lags are thought to be the result of fluctuations originating at large radii and propagating inward with the accretion flow at the inflow timescale \citep[e.g.][]{Davis_2020},
\begin{equation}
    \label{eq:t_inr}
    \tau_{\rm in} = \frac{1}{\alpha \Omega}\frac{r^2}{h^2},
\end{equation}
where $\alpha$ is the \citet{ShakuraSunyaev1973} stress parameter, $\Omega=(GM/r^3)^{1/2}$ is the orbital frequency, $M$ is the mass of the SMBH, $G$ is the gravitational constant, $h$ is the scale height of the disk, and $r$ is the radial position in the disk.

To derive the timescale of the long lag, we convert  $\tau_{\rm in}$, to the inflow velocity, $v_{\rm in}=r/\tau_{\rm in}=\alpha \Omega h^2/r$, and integrate between the radius corresponding to band $x$, $r_x$, and the reference radius, $r_0$,
\begin{equation}
\begin{split}
    \label{eq:dr/v}
    \tau_{\rm lag} = \int_{r_0}^{r_x} \frac{dr}{v_{\rm in}} = \frac{r_0^{2\beta}}{h_0^2 \alpha \sqrt{GM}} \int_{r_0}^{r_x} r^{5/2 -2\beta} dr \\ =\frac{r_0^{2\beta}}{(7/2-2\beta)h_0^2 \alpha \sqrt{GM}} \left[r_x^{7/2-2\beta}-r_0^{7/2-2\beta}\right],
    \end{split}
\end{equation}
where we have added an $r$ dependence to $h$, by setting $h(r)=h_0\left(r/r_0\right)^{\beta}$. This choice is informed by the fact that, even though the ratio of $h/r$ is constant in the classic thin disk model, numerical simulations suggest that the scale height possesses a radial dependence \citep{JiangBlaes2020}. This function form is also more general as a constant scale height is just the special case with $\beta=0$. Without assuming a radial temperature dependence, putting Equation \ref{eq:dr/v} in terms of $\lambda$ gives,
 \begin{equation}
    \tau \propto \left[\left(\frac{\lambda}{\lambda_0}\right)^{\alpha(7/2-2\beta)}-1\right],
 \end{equation}
where $r \propto \lambda^{\alpha}$.

With this relationship, we can use the fitting result $\gamma = -4.96 \pm 1.49$ from Equation 10 to derive the value of $\beta$. If we assume the radial temperature dependence of the \citet{ShakuraSunyaev1973} thin disk model, $T_{\rm eff}^4 \propto 1/r^3$, and use the characteristic emission wavelength $T \propto 1/\lambda$, we get $r\propto \lambda^{4/3}$. Following Equation 13, when keeping $\alpha$ fixed at 4/3, we obtain $\beta = 3.59 \pm 1.22$. Given the relationship between the scale height and radius, this slope implies a scaling relationship where the disk puffs up at larger radii. This interpretation of a puffed-up geometrically thick disk contradicts the traditional picture of a thin accretion disk, but is often invoked to explain the line driving broad absorption line quasars \citep[e.g.][]{Murray1995,Leighly2004}. The best-fitting parameters for all models discussed are shown in Table \ref{tab:results}.

Here, we have not explored the possibility of a long lag extending to the X-ray bands in this work. Given the accretion rate of Fairall 9, we expect that the X-ray emission emerges from only a few to ten gravitational radii and is likely optically thin \citep{Jiangetal2019}. Therefore, the fluctuation propagation that drives the long lag may not translate to the X-ray emission region. Furthermore, the relationship between the X-ray variability and the short lag remains complicated \citep{Edlsonetal2019,MahmoudDone2020,HagenDone2022}, causing us to defer more detailed analysis of the X-ray lightcurves. For future studies, since obtaining more wavelength coverage at higher frequencies is difficult, a more complete demographic analysis of the AGN population will help distinguish if this slope is ubiquitous in all systems. At the same time, numerical simulations of AGN disks can enhance our understanding of the possible distribution of observed lags.

\section{Summary}\label{sec:summary}
Variability studies of AGN not only inform basic properties of their accretion disks, but also have the potential to illuminate the fundamentals of accretion physics. For example, past work such as \citet{Cackett2021} show that multiple lags can be recovered from UV/optical continuum reverberation mapping due to distinct processes occurring in the AGN disk. For this purpose, continuous and high-cadence monitoring of AGN with multi-wavelength coverage, such as the intensive disk-reverberation
campaign of galaxy Fairall 9 is required. We apply both the traditional cross-correlation method with JAVELIN and a Fourier method in frequency space to simulated and real data sets, in an attempt to recover both the well-known and rapidly varying short lag that propagates outward, and the more elusive and slowly varying long lag that propagates inwards, in AGN light curves. 

In our simulated DRW light curves, with a length slightly longer than 2 years and $\sim$day cadence, JAVELIN produces a highly accurate recovery of the long lag while failing to accurately localize the short lag without detrending. The Fourier method can easily identify the timescale at which both processes are occurring, and can very accurately recover the shorter lag while constraining both lags simultaneously. However, JAVELIN short lags and Fourier long lags can be biased by the other lag, especially when the other lag has a strong response. On the other hand, long lags with JAVELIN and short lags with Fourier methods can almost always be accurately recovered. Note that all our tests are conducted with DRW light curves to simulate aperiodic AGN variability. However, discrepancies between this assumption and the real light curve behaviour has been observed \citep[e.g.][]{Mushotzky2011,Kasliwal2015, Smith2018}.

When performing the same analysis on real Fairall 9 light curves, both methods successfully identify short lags that increase in length with respect to the wavelength of observation; the Fourier method measures the short lags with significantly higher precision. Detrended JAVELIN short lags are twice as long as they are from the full undetrended light curves, and better match the \citet{F92020} results. We also present the first recovery of a long lag in AGN measured through cross-correlation with JAVELIN that constrains the long lag at $\sim -70$ days between bands with the largest wavelength difference, propagating inwards in the accretion disk. However, due to the fact that the total time span of the data is less than a year, the frequency-space range is too short for the Fourier method to correctly measure the long lag, especially when the lag becomes much longer between bands with greater wavelength differences. As a result, future work is advised to use the Fourier method when the lengths of available high-cadence light curves are a few times longer than that of the longest lag. With a longer time span, the Fourier method performs significantly better, and can also reveal information that the cross-correlation method is incapable of, such as separating the timescales of distinct processes. A good opportunity for this test is with the upcoming release of the full 1.7-year intensive disk reverberation mapping of Fairall 9 with daily \textit{Swift} and LCO monitoring. Having twice the length of the data we are currently using, this release could probe a larger frequency range with the Fourier method. Additionally, we suggest applying both methods to the same data set, so that the Fourier method can provide verification of recoveries and separate them from spurious signals in the cross-correlation method.

Overall, our methods show promising results that raise the exciting prospect of finding long lags in more AGN, and thus mapping accretion disk structure as a function of AGN parameters. Many other past and future observations can be potentially useful for the same purpose. \citet{Stone2022} compiled \textit{gri} light curves of a sample of 190 quasars within SDSS Stripe 82 spanning $\sim 1998 - 2020$ from SDSS, PanSTARRS-1, the Dark Energy Survey, and follow-up monitoring with Blanco 4m/DECam. Other potentially useful data sets include the ASAS-SN survey in \citet{Yuk2022} and the upcoming decade-long Vera C. Rubin Observatory's Legacy Survey of Space and Time (LSST). Extracting the long lag from these data sets will require accurate analysis since the limited wavelength difference (e.g. from \textit{u} to \textit{i} in LSST) will lead to much shorter long lags. An added challenge here is, in some cases, the sparse sampling or large seasonal gaps. Because of LSST's uniquely high cadence and long time baseline, it will be suitable for searching for the long lag in a larger population of AGN. We will explore this possibility with a suite of simulated LSST light curves in a follow-up paper, with improved metrics to overcome inevitable seasonal gaps (Secunda et al. in prep). Future work can focus on distinguishing spurious signals from these sources in both frequency and temporal space, which may aid lag recovery. Additionally, testing our methods on a select sample of the these quasars, especially high-redshift galaxies where lags will become more prominent even with short wavelength coverage, will further verify the applicability of both the Fourier method and JAVELIN overall. The derived lag distribution from such a population of AGN will in turn deliver precious information regarding accretion physics from distant galactic central engines. 

\section*{Acknowledgements}
We thank A. J. Barth for helpful feedback. The Center for Computational Astrophysics at the Flatiron Institute is supported by the Simons Foundation. Part of the work was done when YFJ was attending the Binary22 program in KITP, which was supported in part by the National Science Foundation under Grant No. NSF PHY-1748958. A.S. is supported by a fellowship from the NSF Graduate Research Fellowship Program under Grant No. DGE-1656466. JEG is supported in part by NSF grants AST1007052 and AST1007094.

\bibliography{ref}{}
\bibliographystyle{aasjournal}

\appendix

\section{Simulated light curves}\label{app:sim}

    In this section, we test both JAVELIN and the Fourier method with a variety of DRW light curves covering a range of parameters. Hence, we demonstrate here that the recovery of short and long lags in Section \ref{sec:F9} is robust and reasonable. Given its sampling cadence and wavelength coverage, the intensive disk-reverberation mapping campaign of Fairall 9 is one of the most unique and extensive datasets currently in existence. This is also an improving set of light curves which is expected to double its current length in the foreseeable future. As a result, many of the test cases presented in this paper are built from the characteristics of the Fairall 9 data set, including the cadence, length, and error of our simulated light curves.
    \subsection{Simulations} To demonstrate the ability to constrain lags for each of our methods, we generate driving light curves following \citet{Timmer1995}'s DRW model assuming a PSD with a slope of $D(\nu)\propto \nu^{-2}$. Then, we reprocess the light curve in frequency space with a response function following equation \ref{eq3}; when necessary, we apply an inverse-Fourier transform to $R(\nu)$ to obtain the reprocessed light curve in temporal space. Since we expect lags both with $\sim$ several-day and $\sim$ 100-day timescales in opposite directions, we construct our response function as a composite of two Gaussian functions each in the form of:
    \begin{equation}\label{eq8}
    \rm \psi(t) = \frac{1}{St\sqrt{2\pi}}exp^{\frac{-(t-M)^2}{2S^2}}.
    \end{equation}   
    We then vary M and S to change the amplitude and width ratio between the two Gaussian functions. Unless otherwise specified, we use an amplitude ratio of 5:1 and a width ratio of 1:24 between the response functions of the short and long lags. We place our Gaussian centroids at +5 and -120 days respectively for test purposes such that they take place at different timescales and opposite directions, thus matching the physical motivation discussed earlier. The resultant response function is presented in the first panel of figure \ref{fig:uniform} with a range of amplitude ratios from 1 to 0.01 between the -120-day and +5-day Gaussian functions. To create lags with opposite directions, the negative lag is a Gaussian function with negative time. Intuitively, larger S means a broader distribution, and M is the mean of the Gaussian. 
    
    For uniform-cadence light curves, given current limitations of ground-based observatories, we adopt a one day cadence and a 1 year length that matches the best currently available light curves. For non-uniform cadence light curves, we regenerate the DRW light curve and sample the data points according to the Modified Julian Dates (MJD) in the data from \citet{F92020}. We use the signal-to-noise ratio of the same data sets to generate errors and perturb our simulated light curves. 
    
    \subsubsection{Uniform cadence time-series}
\begin{figure*}[ht]
        \centering
        \includegraphics[width=\textwidth]{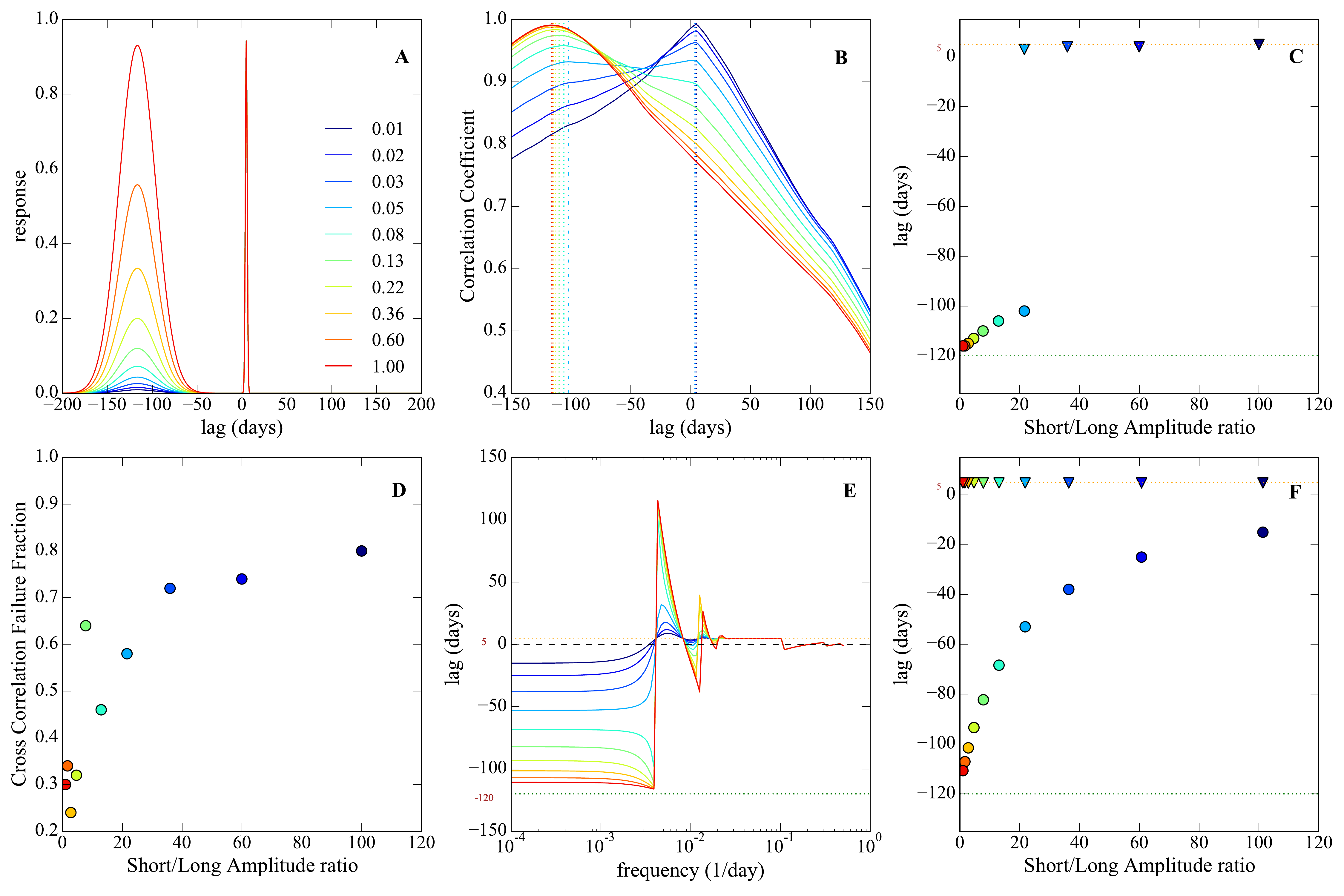}
        \caption{Comparison between the frequency-resolved and cross-correlation approaches as a function of the amplitude ratio between the two lags' response functions. Panel A show the concatenated response function with a varying long lag amplitude and constant short lag response. Panels B and C show the cross-correlation function and the recovered peaks in the functions respectively. Panel D demonstrates that cross-correlation can identify both lags easily when the amplitude of the response functions are comparable, with the failure rate increasing with diminishing long lag amplitude. Panels E and F show the recovered lag in frequency space and as a function of amplitude ratios respectively. We observe that as the long lag becomes less obvious, the recovered lag may not be an accurate determination of the input lag.}
        \label{fig:uniform}
    \end{figure*}

    With uniformly-sampled light curves, we are able to use both the simple cross-correlation analysis presented in \ref{sec:CC} and simple frequency-resolved analysis presented in \ref{sec:freq}. As seen in figure \ref{fig:uniform}, the cross-correlation method identifies the long lag easily when its amplitude in the response function is comparable to its shorter lag counterpart, and the short lag is detectable at most ratios except when the amplitude of the long and short lags are comparable. On the other hand, frequency-resolved analysis can recover both lags at all frequencies, with outstanding accuracy and consistency for the short lag, but the accuracy for the long lag decays with the strength of its signal. We should also note that the cross-correlation method can fail occasionally when asked to constrain both the long and short lags simultaneously. By running 100 test runs of cross-correlation on random simulated light curves, the fraction that fails to identify both lags increases with decreasing signal strength from the longer lag. The cross-correlation example in figure \ref{fig:uniform} further elaborates on this by showing the inability for cross-correlation to identify the long lag when the amplitude ratio falls below 5\%.
    
    \begin{figure*}[ht]
        \centering
            \includegraphics[width=\textwidth]{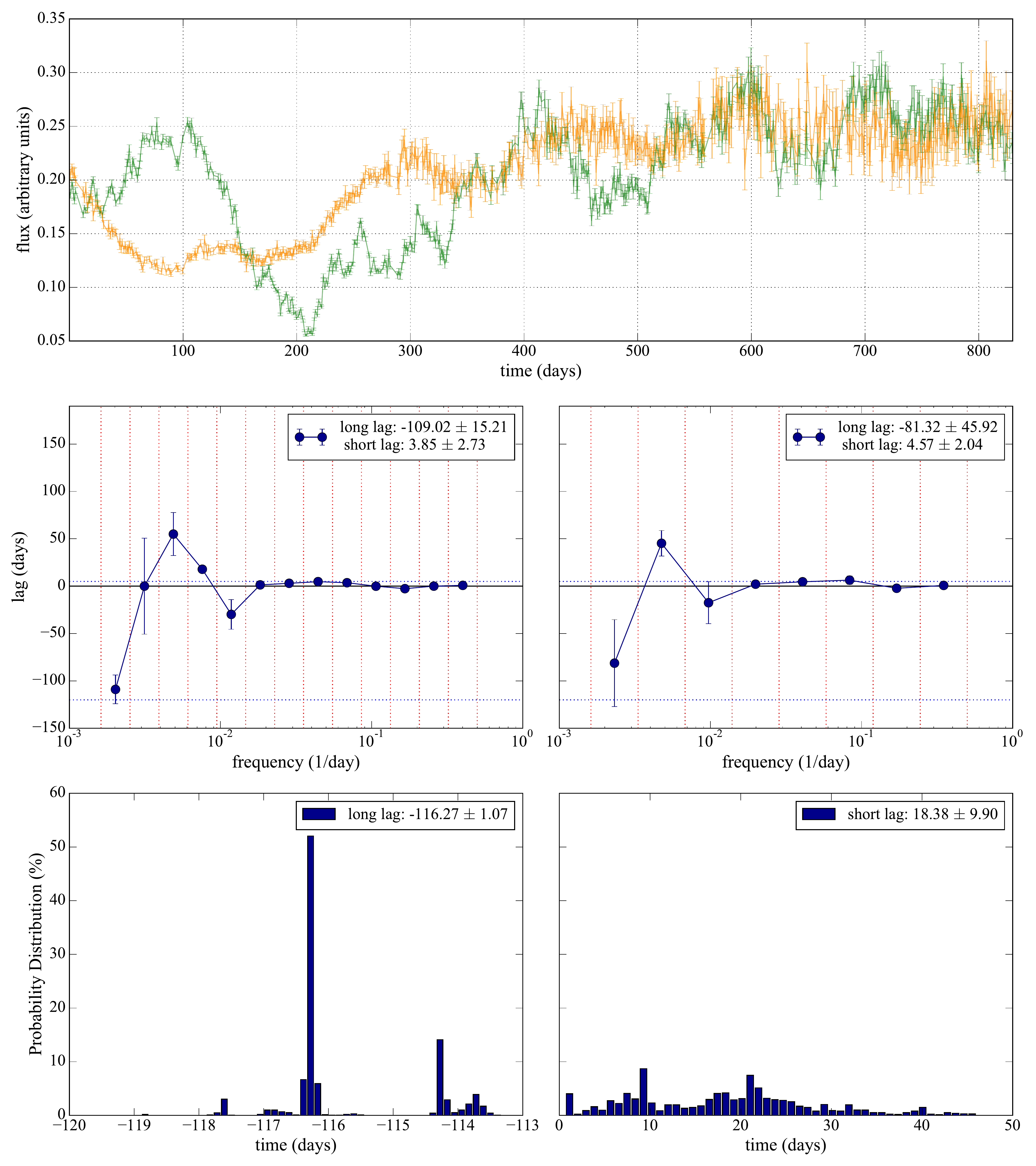}
        \caption{Comparison between the Fourier and JAVELIN lag recoveries. The top panel shows the simulated driving (green) and reprocessed (orange) light curves. The middle panels show the recovered lags by the Fourier method with two choices of binning (13 bins and 8 bins), which can have some impact on the accuracy of lag determination. The bottom panels show the recovered long (left) and short (right) lags by JAVELIN, where interpolation might have hindered a precise measurement of the short lag. }
        \label{fig:zog1}
    \end{figure*}

    \begin{figure*}[ht]
        \centering
        \begin{tabular}{c}
            \includegraphics[width=\textwidth]{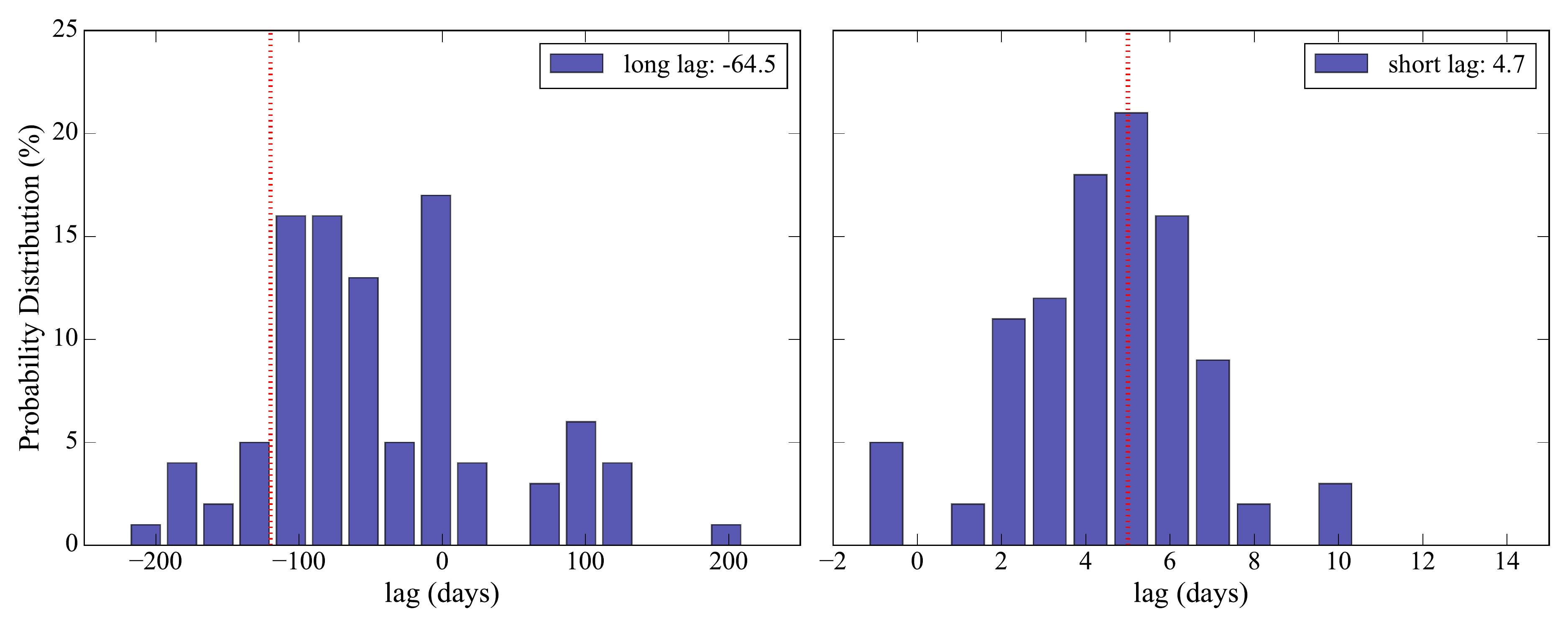}\\
            \includegraphics[width=\textwidth]{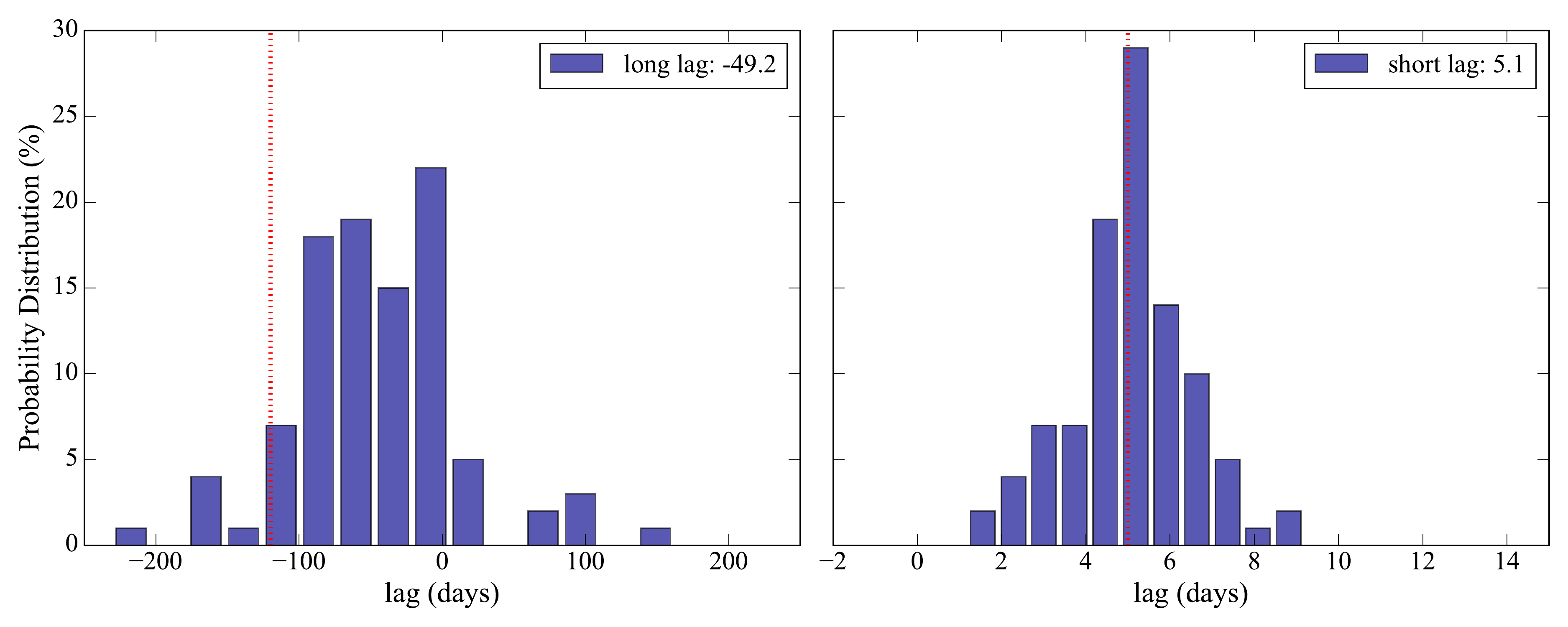}\\
            \includegraphics[width=\textwidth]{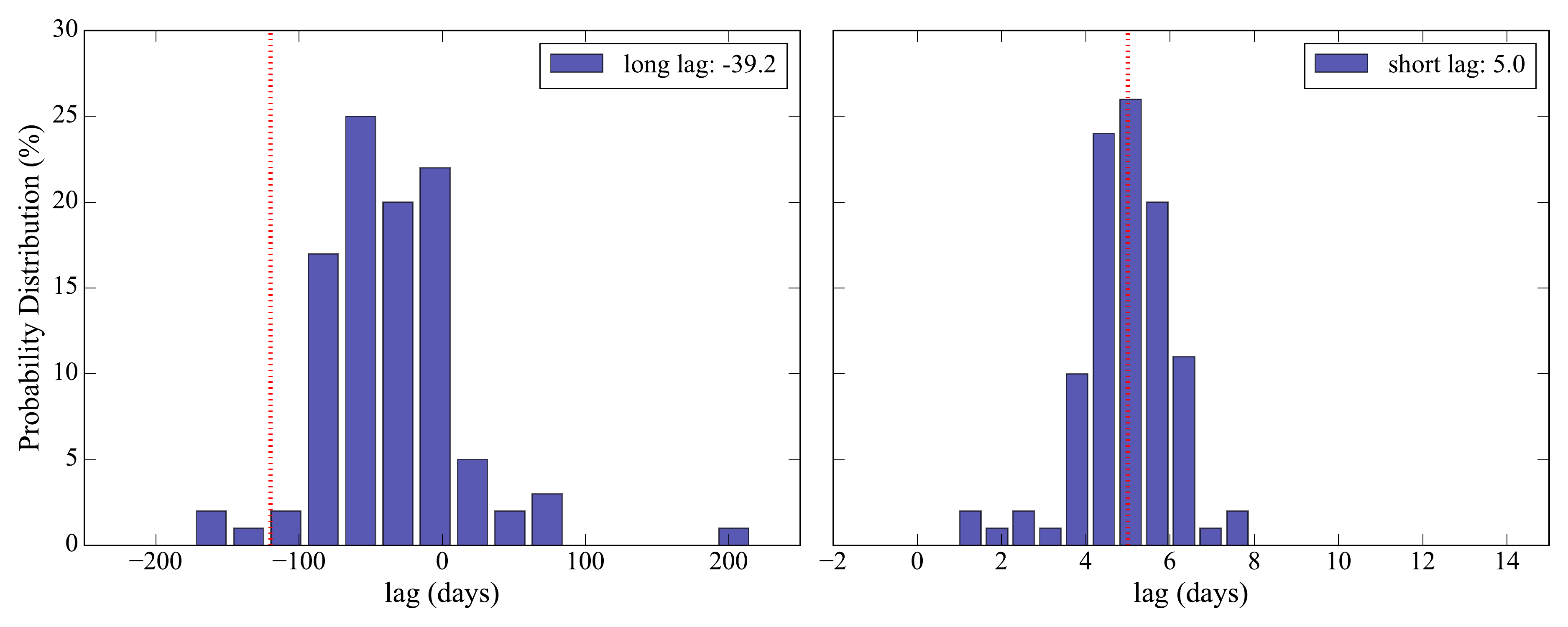}\\
        \end{tabular}
        \caption{Results of 300 simulated light curves analyzed via the Fourier method. The top, middle, and bottom panels each represents 100 light curves reprocessed with response functions where the amplitude of the long lag is 20\%, 10\%, and 5\% of that of the short lag respectively. The legend shows a median value of the recovered lags, and the red dotted lines are the input lag at -120 and +5 days. 78\%, 85\%, and 89\% of the short lags recovered are significant ($>3\sigma$ from 0). We see a clear trend where the median value approaches the input lag when the long lag signal is more prominent. In all three cases, the short lag can be easily recovered.}
        \label{fig:res_amp}
    \end{figure*}

    \begin{figure*}[ht]
        \centering
        \begin{tabular}{c}
            \includegraphics[width=\textwidth]{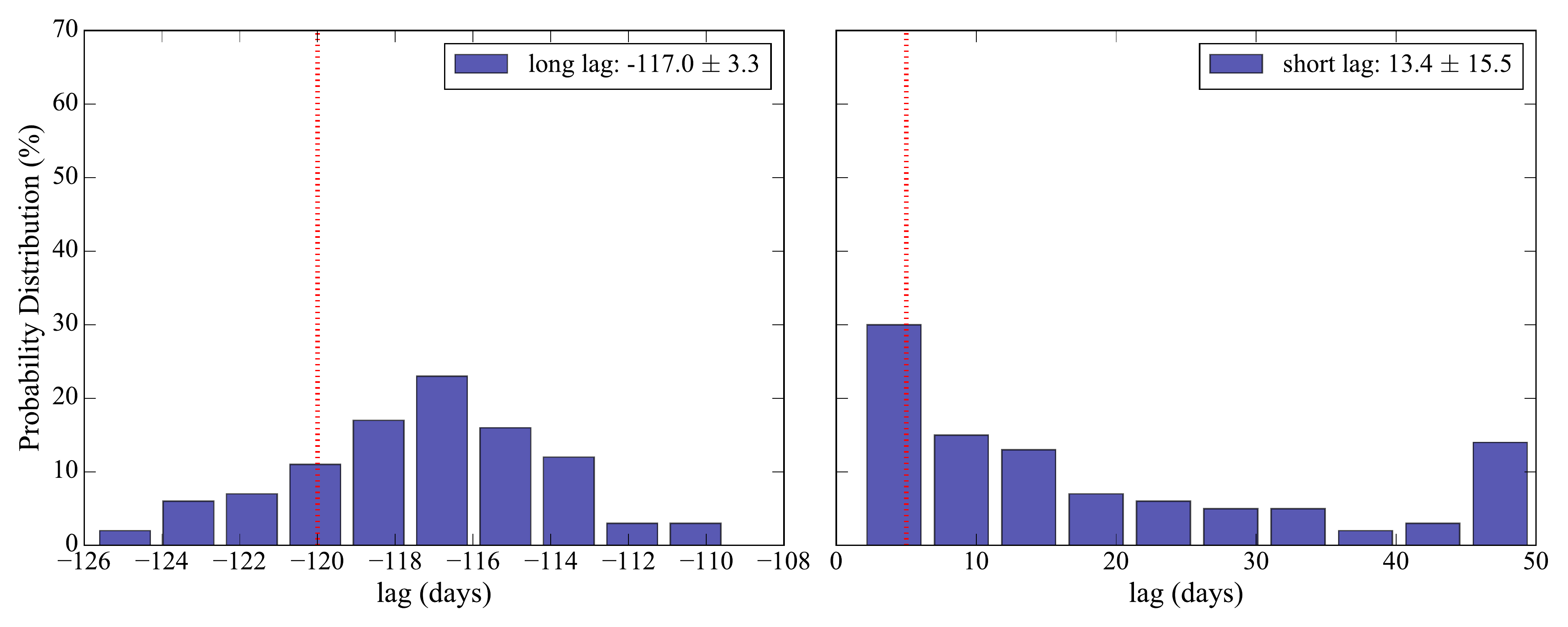}\\
            \includegraphics[width=\textwidth]{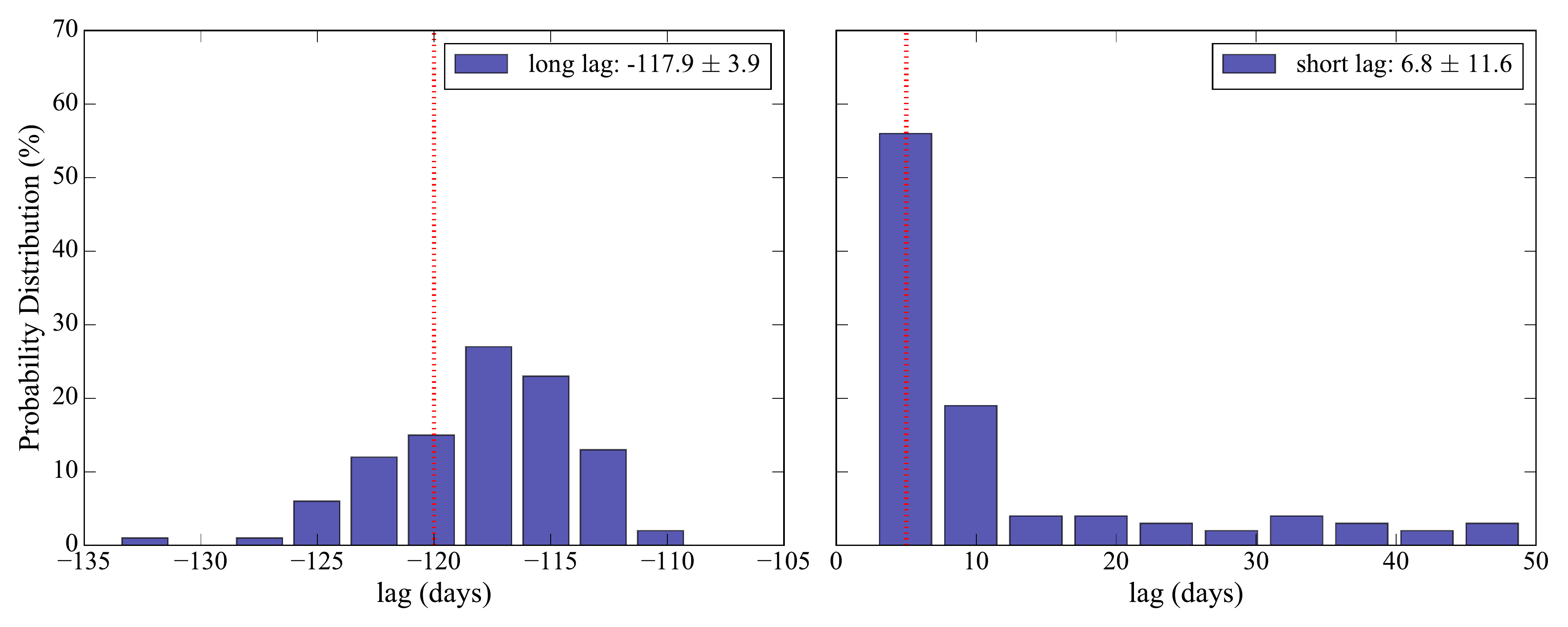}\\
            \includegraphics[width=\textwidth]{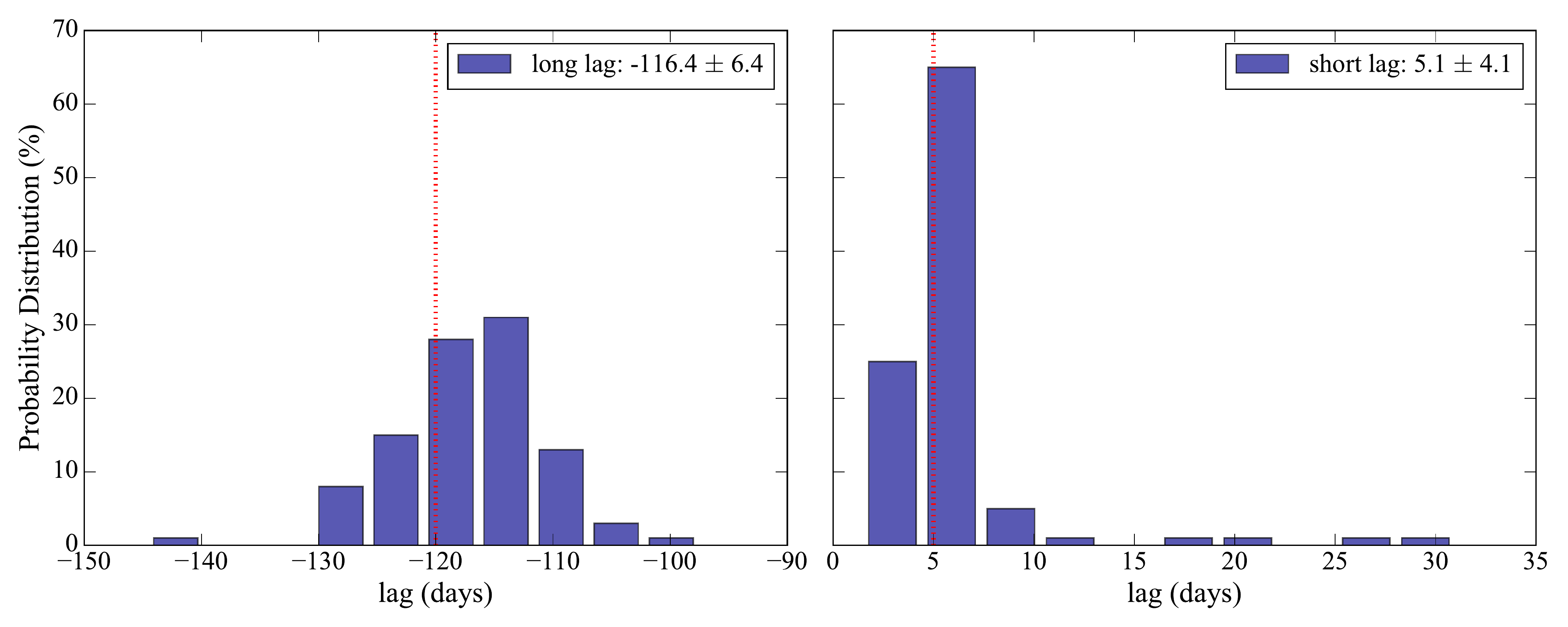}\\
        \end{tabular}
        \caption{Results of 300 simulated light curves analyzed via JAVELIN. The top, middle, and bottom panels each represents 100 light curves reprocessed with response functions where the amplitude of the long lag is 20\%, 10\%, and 5\% of that of the short lag respectively. The legend shows a median value of the recovered lags, and the red dotted lines are the input lag at -120 and +5 days.}
        \label{fig:res_amp_JAV}
    \end{figure*}
    
    \subsubsection{Non-uniform cadence time-series}
    Since UV/optical variations arise at a timescale much longer than a day, astrophysical observations naturally produce gaps in their time-series, hence obtaining uniformly sampled data becomes infeasible. JAVELIN and the \citet{Zoghbi2013} methods act as alternatives in this regime for lag determination without significantly compromising the results of the analysis. 
    
    The driving light curve in Figure \ref{fig:zog1} comes from the DRW model with a length of 829.86 days. This length is triple that of the $UVW2$ (far-UV from \emph{Switft}) band light curve of \citet{F92020}; the flux uncertainty comes from the S/N ratio in the Fairall 9 light curve as well. The Fourier method relies on parametrizing values of power and lags in a preset of frequency bins, but the choice of the number of bins can have a significant impact on the lags we can recover. The lowest frequency bin is set by the total length of the light curve. Having too many bins tends to distribute the power to separate bins while significantly increasing the uncertainty of the time lag with biases introduced. With an appropriate number of bins, the short lag is often visible with negligible errors independent of the light curve we use, usually confirmed by several of its adjacent bins in the frequency range between  $10^{-2}$ and $10^{-1}$. With input lags at +5 and $-120$ days, when adopting 13 bins, $\sim 3$ bins in this range measures an average short lag of $3.85 \pm 2.73$, but with 8 bins, only 2 of them show a short lag. In this case, 13 bins is more suitable because it also returns a more accurate long lag at $-109 \pm 15$ days compared to having 7 bins which returns a long lag of $-81 \pm 46$ days. In virtually all test cases, the long lag tends to lie within 1 sigma of the lag in the first or second frequency bin; with a sufficiently long light curve, we can see a peak that indicates the exact timescale of the long lag. 
    
    To compare the performance of this method in various astrophysical conditions with different physical properties, we show in figure \ref{fig:res_amp} the results of 300 different simulated light curves with the same parameters as above but reprocessed with different response functions: the amplitude ratio between the long lag and the short lag is 20\%, 10\% and 5\% respectively. Because the length of the light curve is not significantly longer than the longest input lag, we take the longest lags in the lowest frequency bins as the long lag. We then take a weighted average for the short lag from bins within frequency ranges between $10^{-2}$ and $10^{-1}$. Each lag is weighted by the inverse square of its error. These bin choices are informed by the fact that these two lags would almost always occur at their characteristic Fourier frequencies; additionally, these bins produce results that are most consistent with the input and cross-correlation lags in our test runs. The Fourier method continues its success in the frequency domain at recovering the short lag in all cases. The recovered long lag, on the other hand, becomes more accurate when its amplitude in the response function is high, and vice versa. However, we note that in about 70\% of all test cases, the Fourier approach can confidently recover a long lag.
    
    By interpolating the same light curve to uniform cadence and to 10 times more data points, JAVELIN is more consistent in recovering a long lag with even higher precision. In this case, the median value of the recovered long lag is $-113.03 \pm 2.68$ days. However, interpolation at short timescales seem to introduce higher uncertainties to measuring the short lag: our input short lag of +5 days lies just outside the 1 sigma region of the recovered short lag at $20.94 \pm 13.02$ days. A lot more scatter is seen in recovering the short lag.
    
    We then apply JAVELIN to the lag extraction of the 300 simulated light curves. Each long and short lag is taken from the median of the probabilistic distribution from each test run. We can observe from Figure \ref{fig:res_amp_JAV} that the recovered long lag is very accurate and precise in all cases. On the other hand, by observing the scatter in the right panel, the recovered short lag becomes much more accurate when the signal from it is much stronger than that from the longer lag.
    
    However, a 800-day high cadence light curve is not always available for targets of interest. Similar tests that we carried out with a $\sim 300$-day light curve (3x shorter) shows that the Fourier method can still accurately and precisely recover the short lag at $5.6 \pm 1.0$ days, but the limited length shortens the frequency range and we cannot recover the full long lag resulting in a measurement of $-25.1 \pm 26.1$ days. On the other hand, JAVELIN still produces a scattered estimate of the short lag at $33.7 \pm 9.2$ days, but is consistent in recovering the long lag at $-120.2 \pm 1.9$ days.  
    
    To confirm that our results do not come from artificial sources such as sample cadence and light curve length, we performed all previous tests on two randomly generated DRW light curves with no reprocessing nor any correlation between them. In these cases, neither method recovers any statistically significant lag.
    
    Results from the two methods allow us to conclude that both methods have their strengths and weaknesses, especially among different physical regimes. For different light curve lengths and cadences, when recovering the long lag, the Fourier method requires the light curves to be a few times longer than the long lag while JAVELIN can provide accurate result with a much shorter one. When recovering the short lag, regardless of the length or cadence, the Fourier method performs much better even lacking an interpolation to uniform cadence. Both method works best when the AGN is sampled with $\sim$day-cadence. One unique strength of the Fourier method JAVELIN lacks is that the former informs the precise timescales at which the physical processes take place. As a result, combining these two methods when analyzing light curves is the best way to constrain two lags with opposite directions simultaneously. To place the tightest constraint on both lags, future work is advised to use the results from the Fourier method as a prior for JAVELIN to carry out cross-correlation studies.

    
    \begin{figure*}[ht]
        \centering
        \begin{tabular}{c}
            \includegraphics[width=0.85\textwidth]{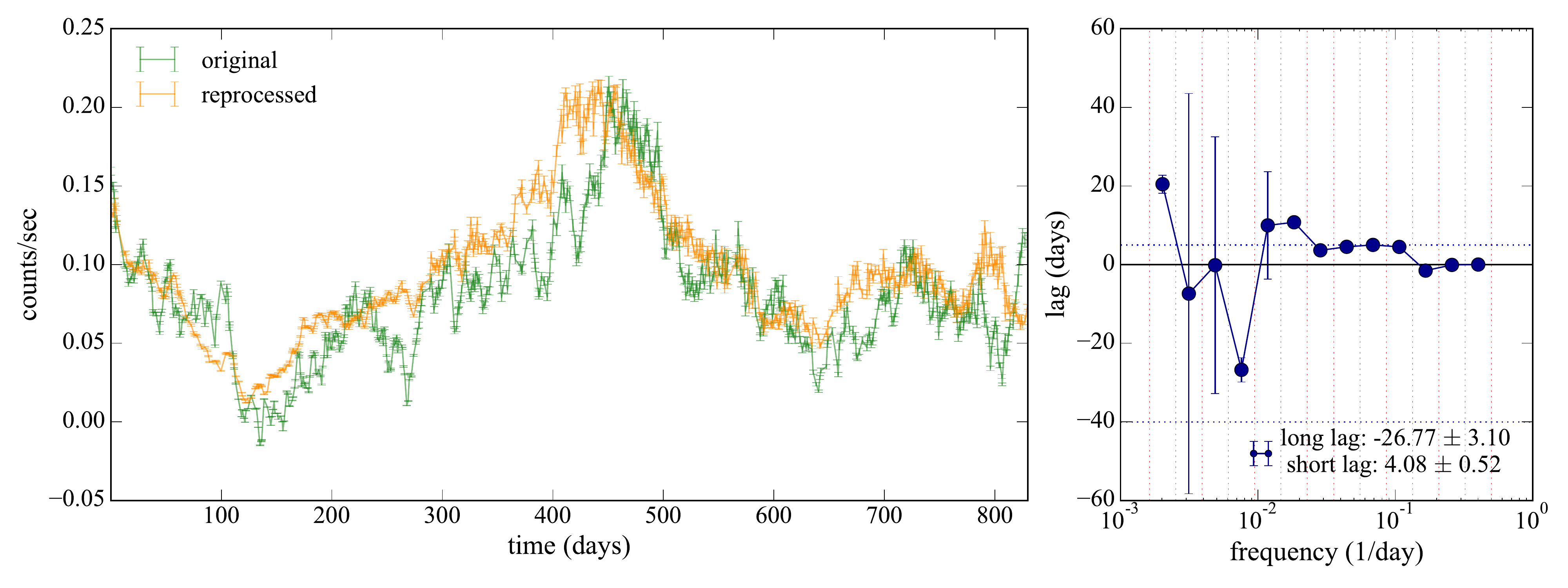}\\
            \includegraphics[width=0.83\textwidth]{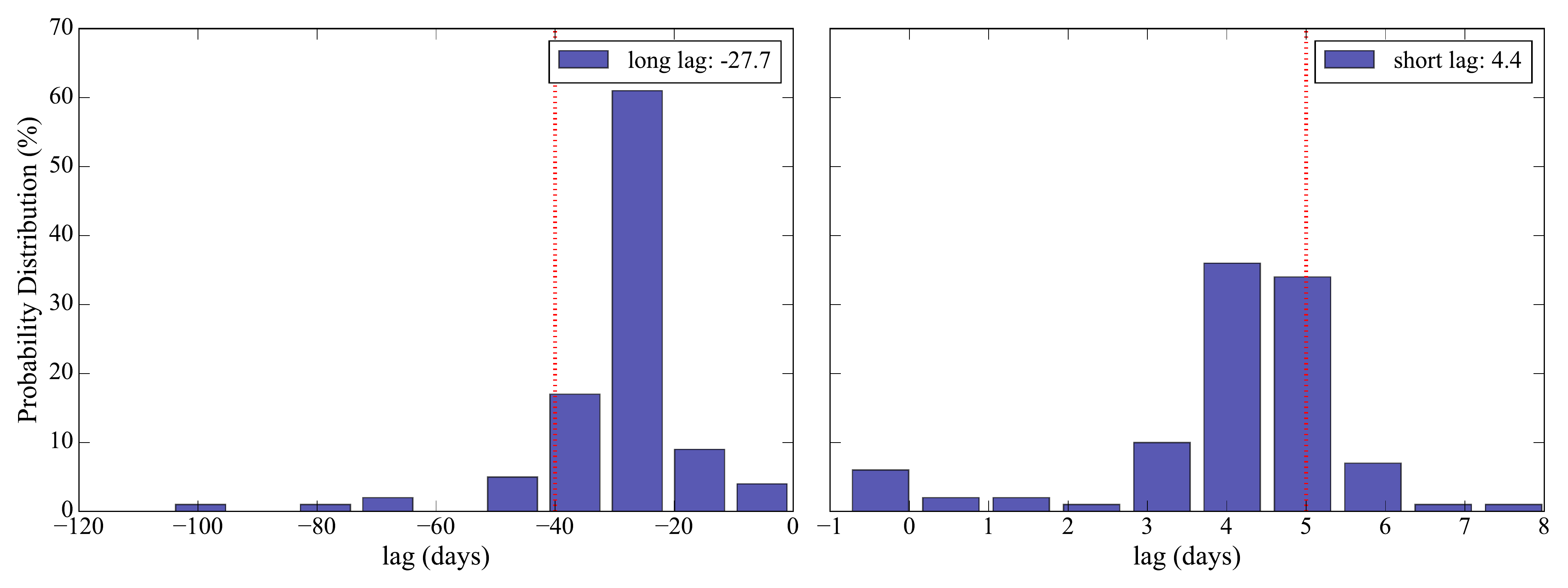}
        \end{tabular}
        \caption{Results of 100 simulated light curves analyzed via the Fourier method. The top panel shows one of the simulations and its resultant lag distribution in frequency space. The input lags are -40 and +5 days with an amplitude ratio of 1:5. 80\% of the short lags recovered are significant ($>3\sigma$ from 0). This test demonstrates the ability of this method to clearly identify the long lag with a peak in frequency space when the length of the light curve is significantly longer than the longest lag.}
        \label{fig:lags_40}
    \end{figure*}
    
    \begin{figure*}[ht]
        \centering
            \includegraphics[width=0.9\textwidth]{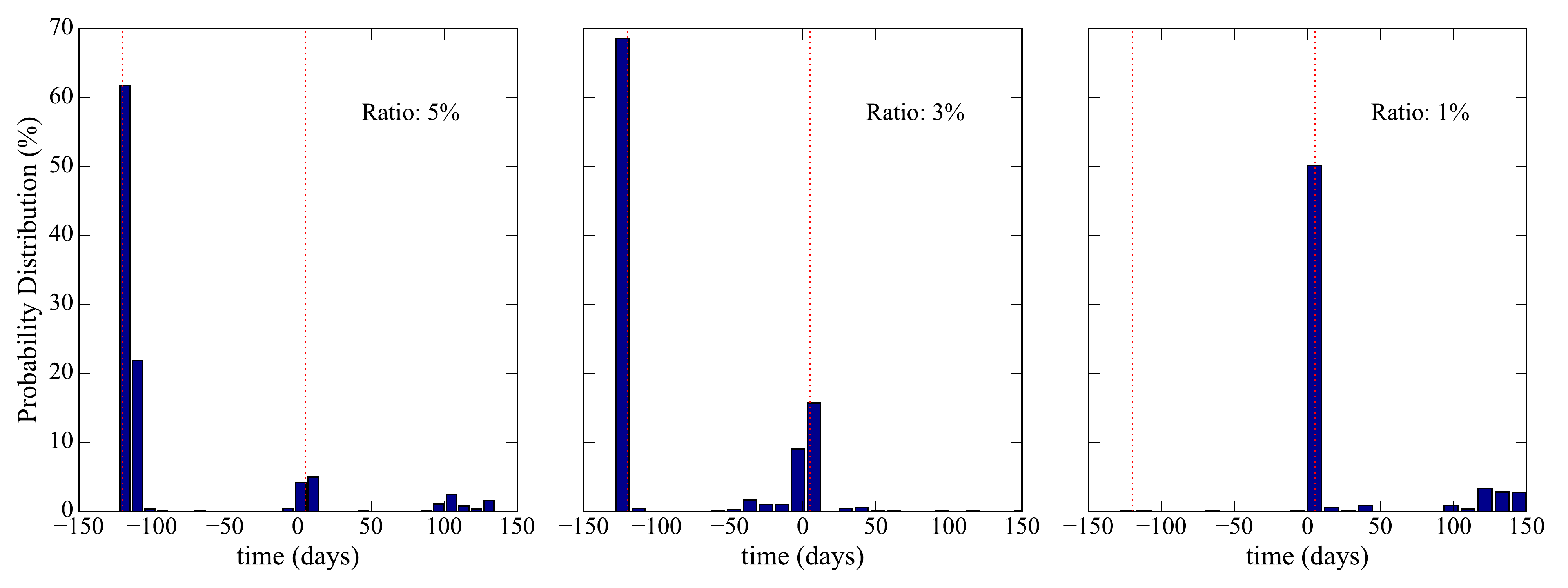}
            \caption{The cross-correlation function from JAVELIN indicating the most probable lags. The plots show results from different input response functions with amplitude ratios between the long lag and the short lag at 5\%, 3\% and 1\% respectively. The red dotted lines show the input lags at -120 and 5 days respectively. The recovered short lag strengthens when the signal from the long lag weakens.}
        \label{fig:ratio_diff}
    \end{figure*}
    
    \begin{figure}[H]
        \begin{tabular}{ccc}
            \includegraphics[width=0.3\columnwidth]{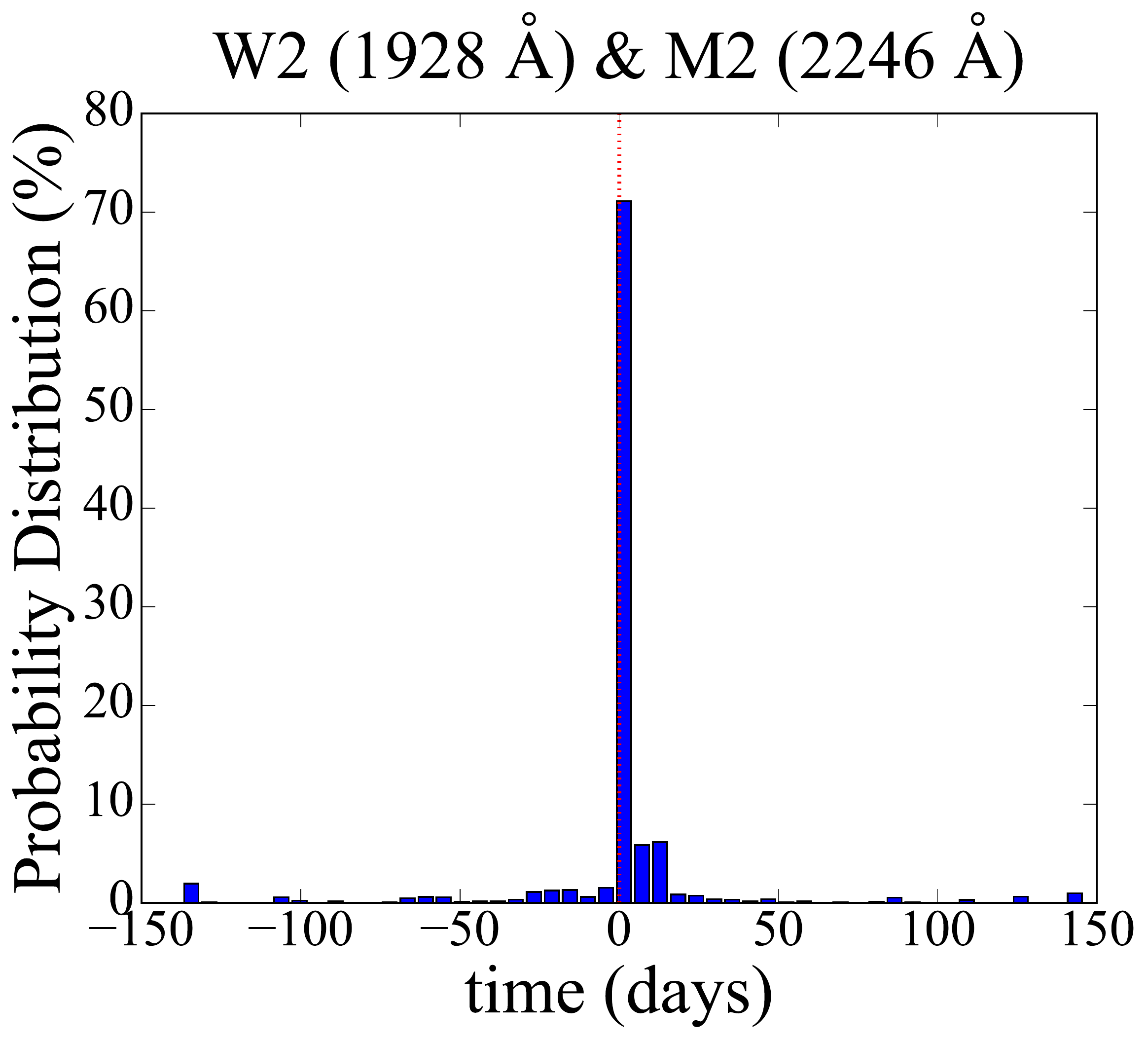} &
            \includegraphics[width=0.3\columnwidth]{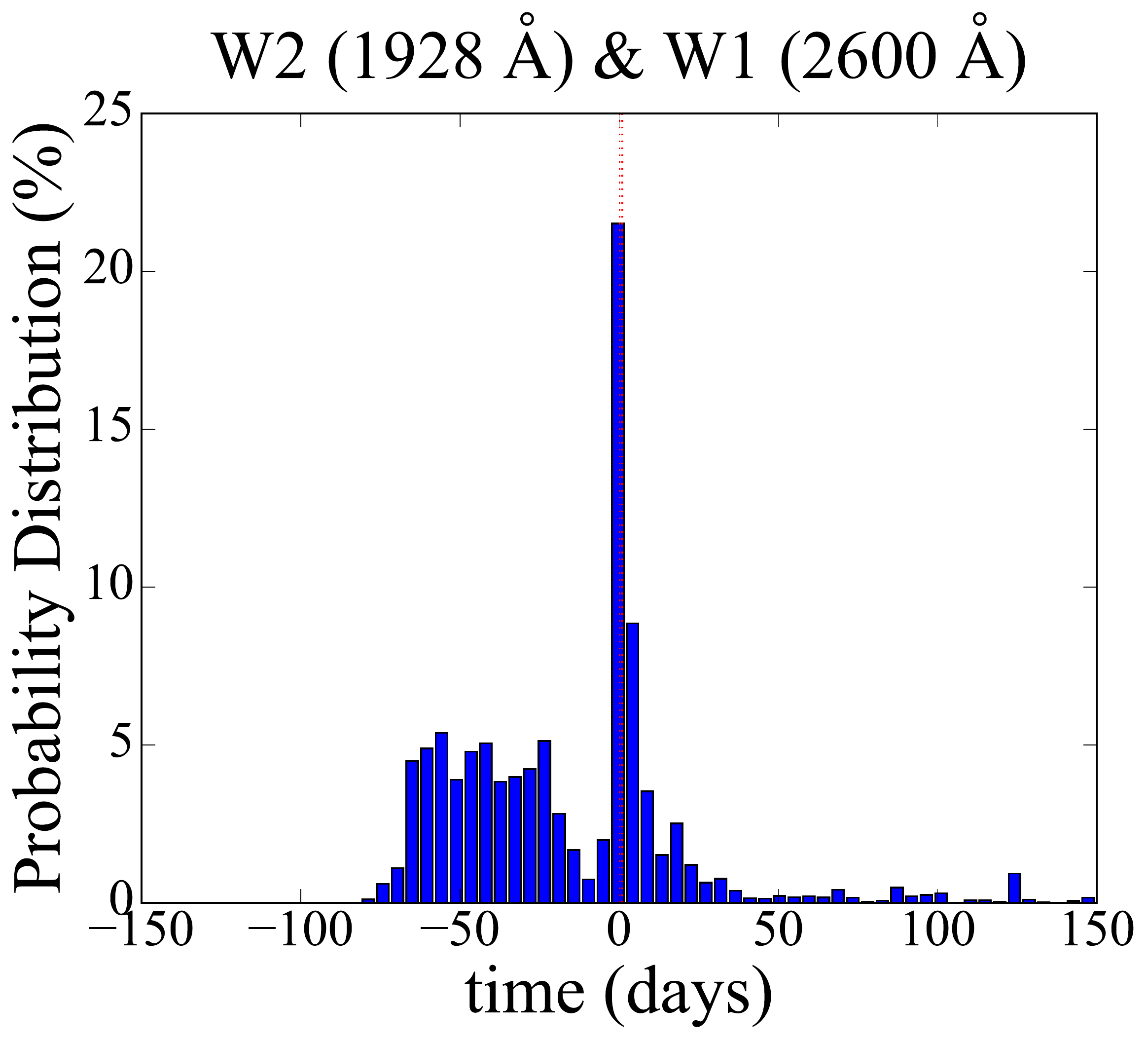} &
            \includegraphics[width=0.3\columnwidth]{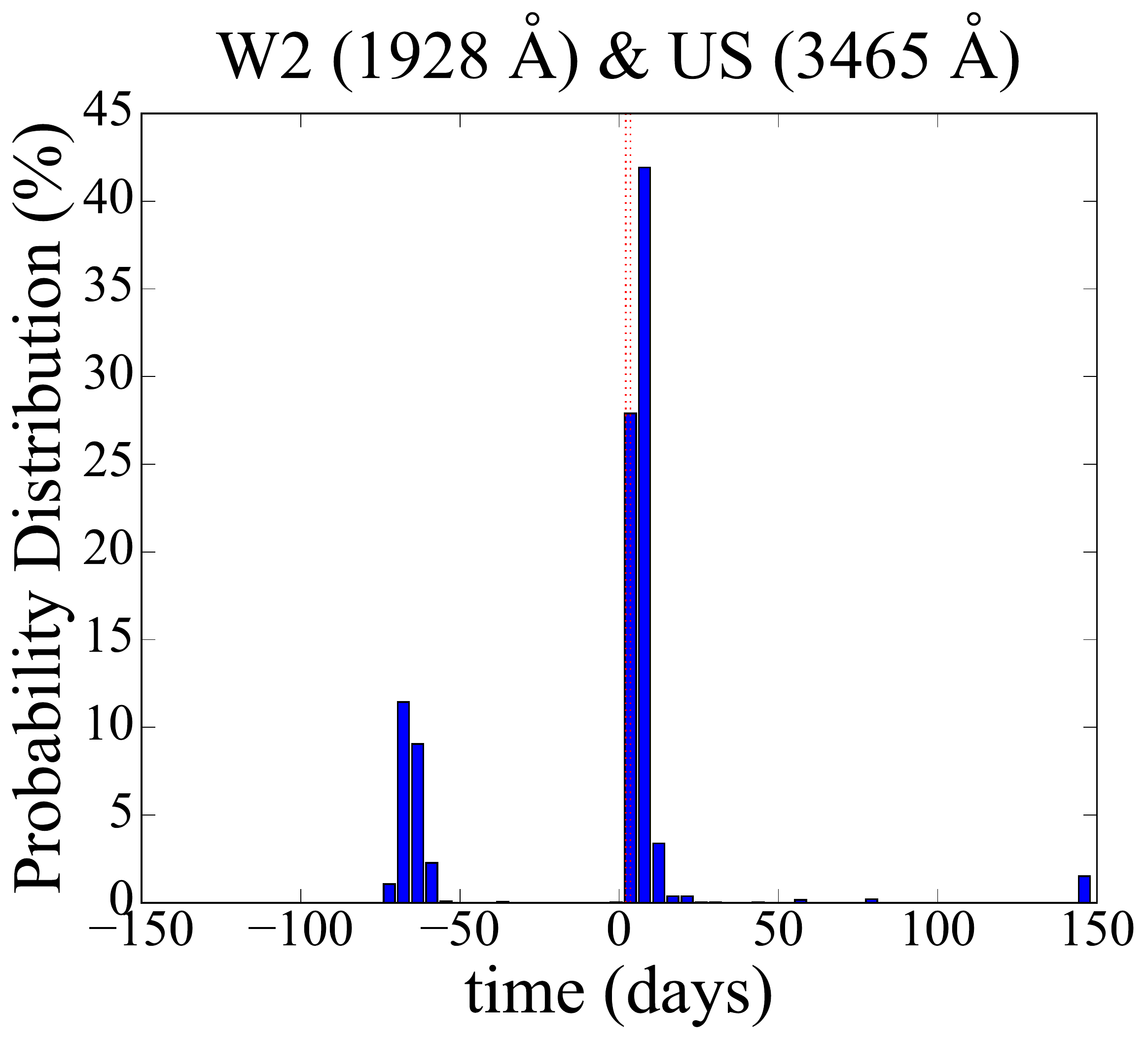} \\
            \includegraphics[width=0.3\columnwidth]{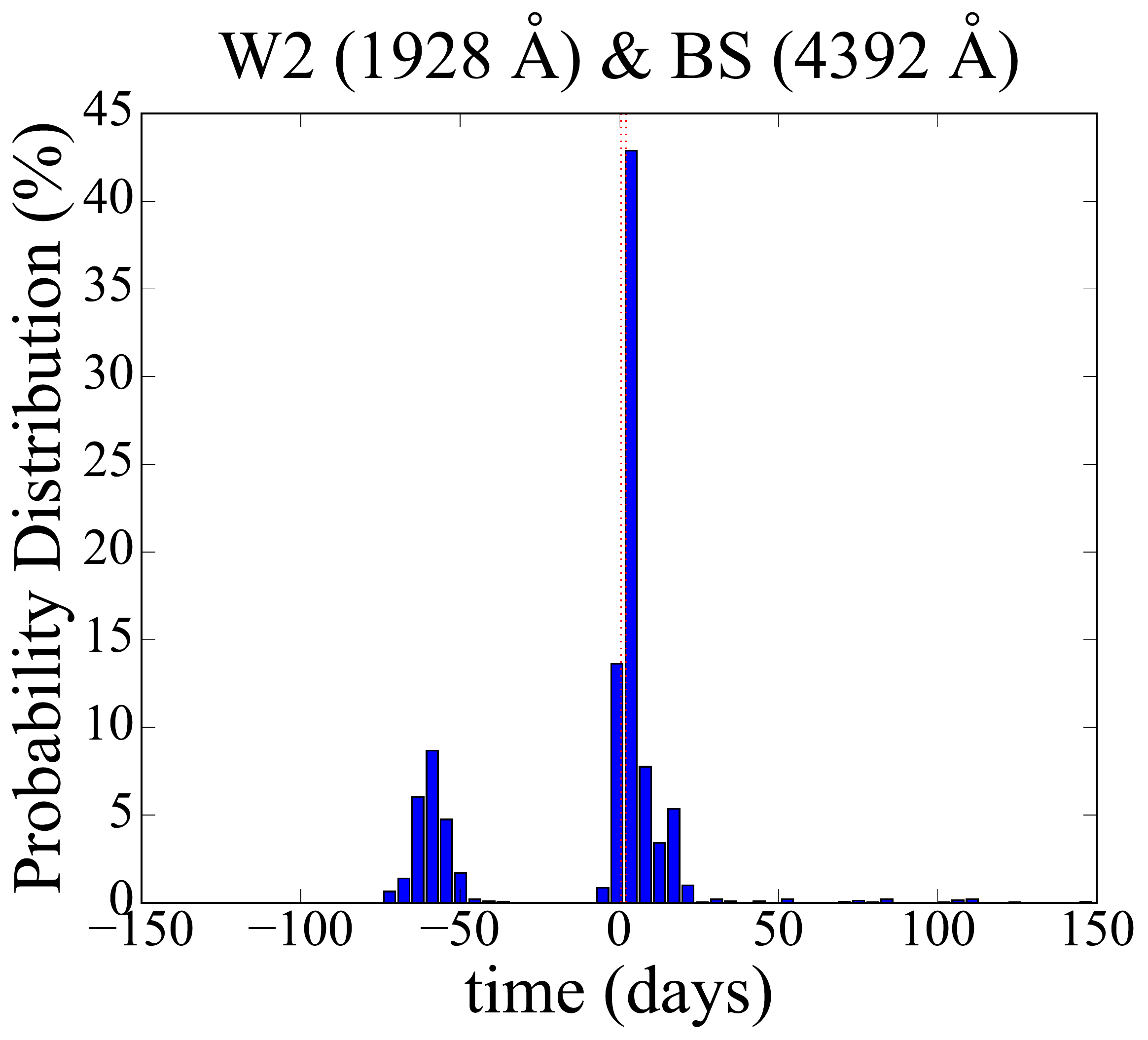} &
            \includegraphics[width=0.3\columnwidth]{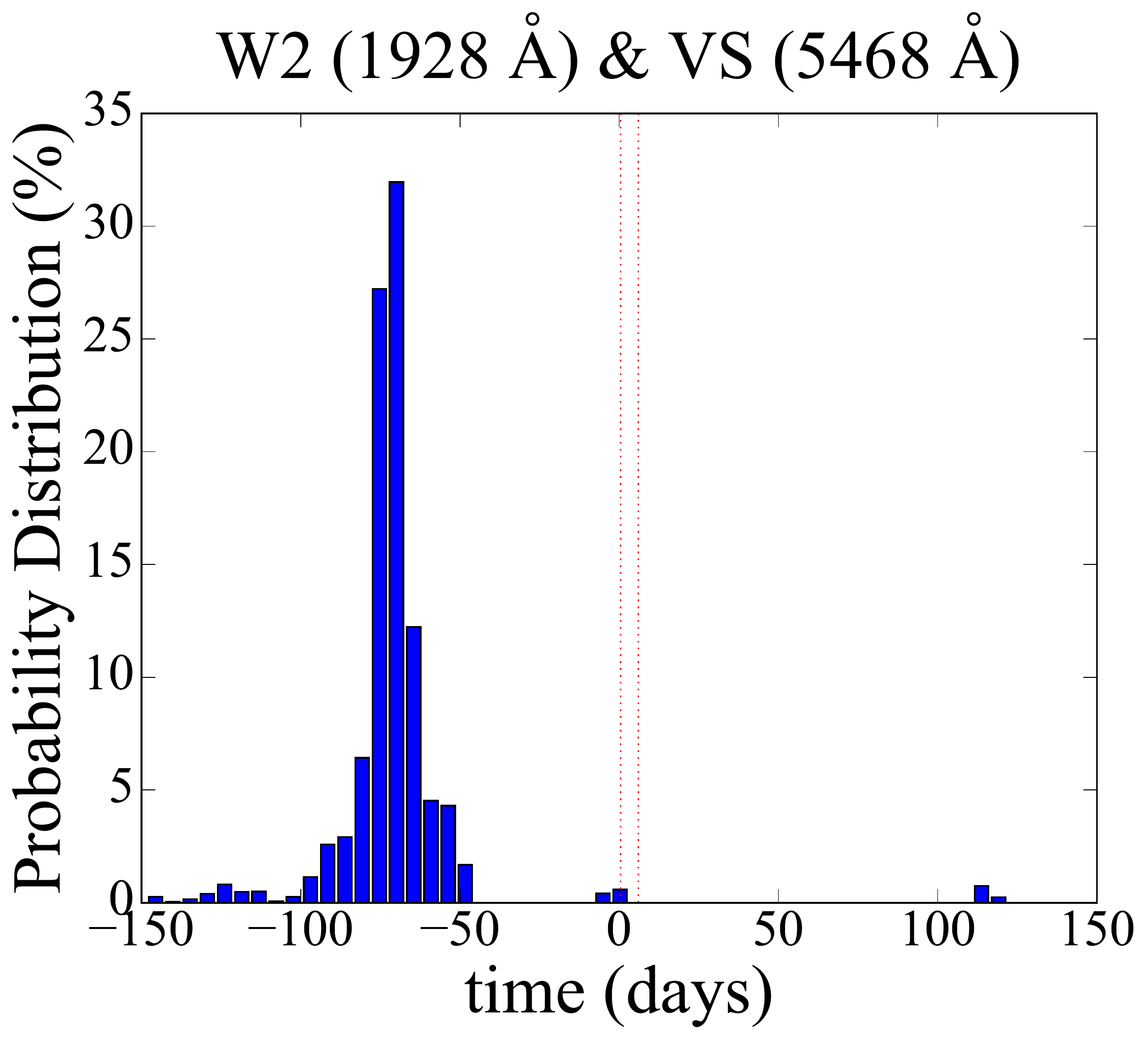} &
            \includegraphics[width=0.3\columnwidth]{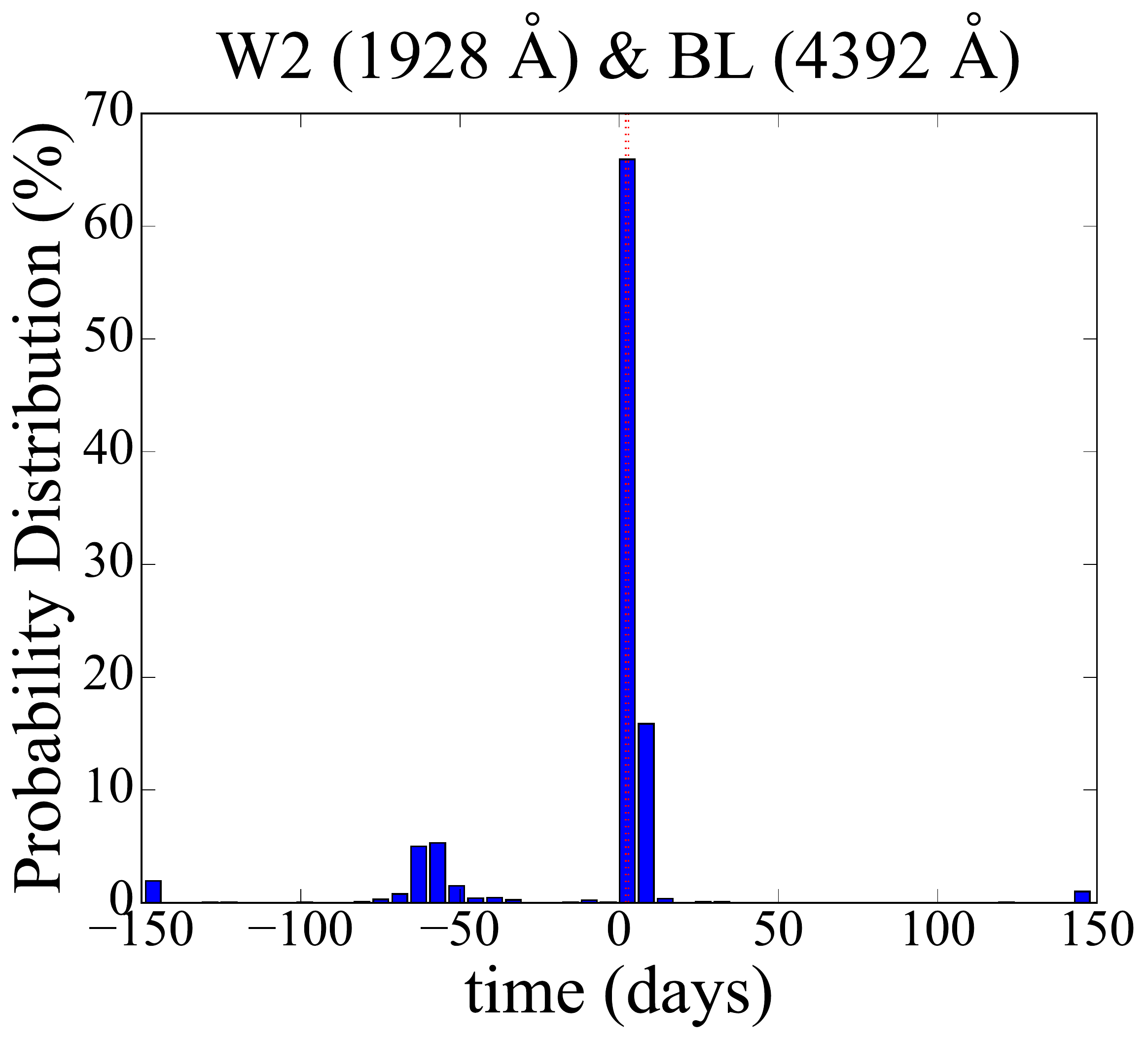} \\
            \includegraphics[width=0.3\columnwidth]{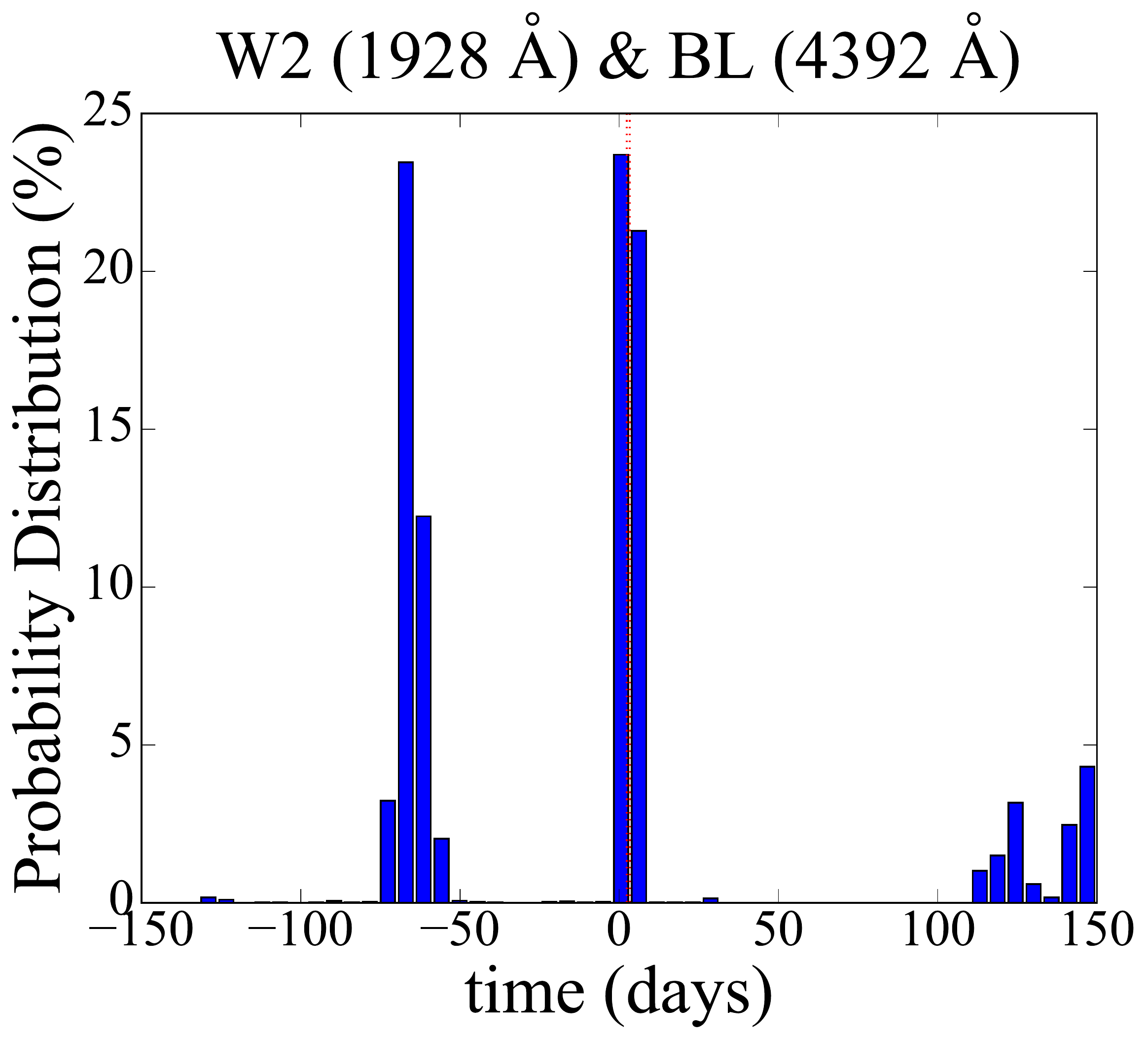} &
            \includegraphics[width=0.3\columnwidth]{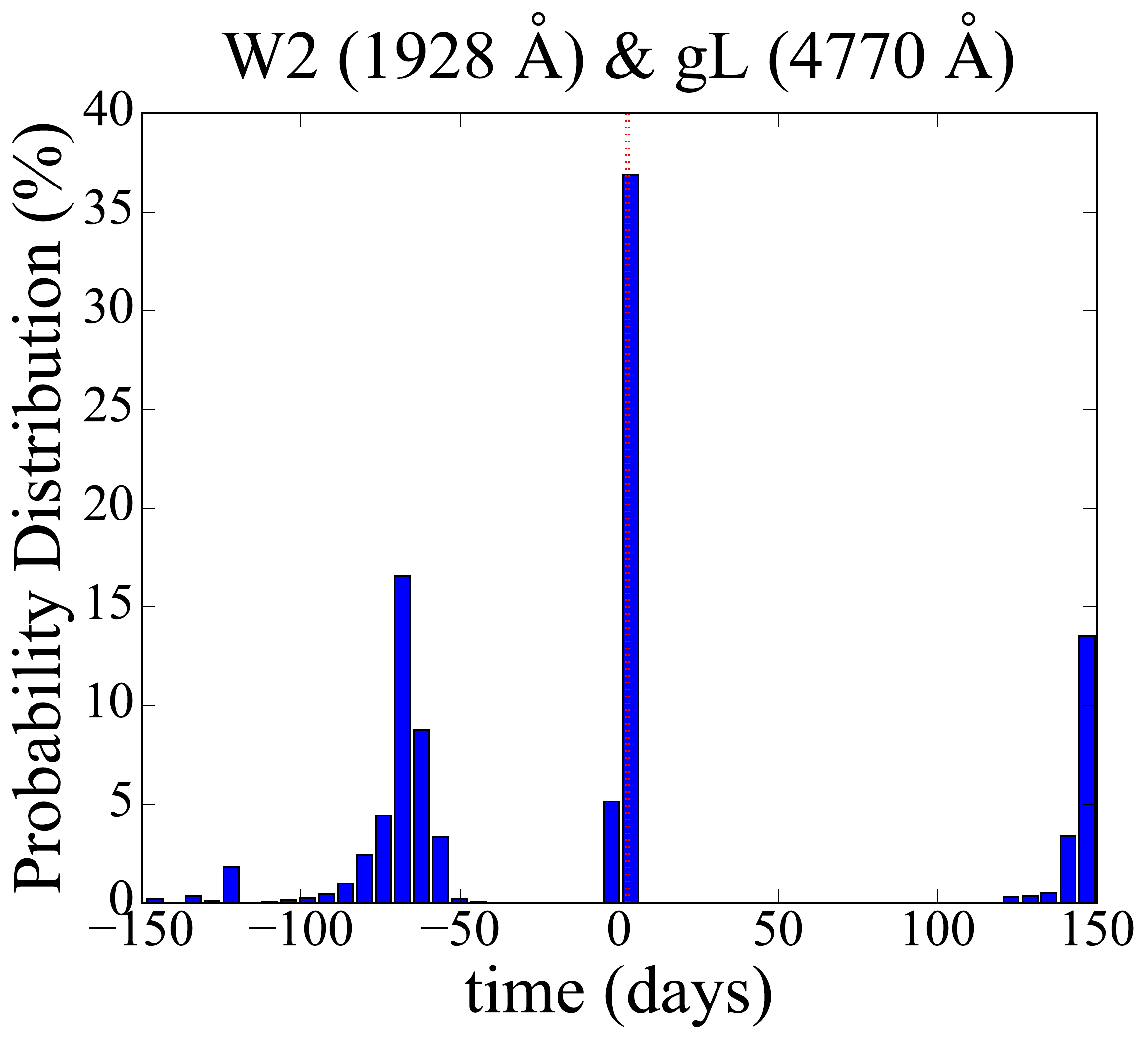} &
            \includegraphics[width=0.3\columnwidth]{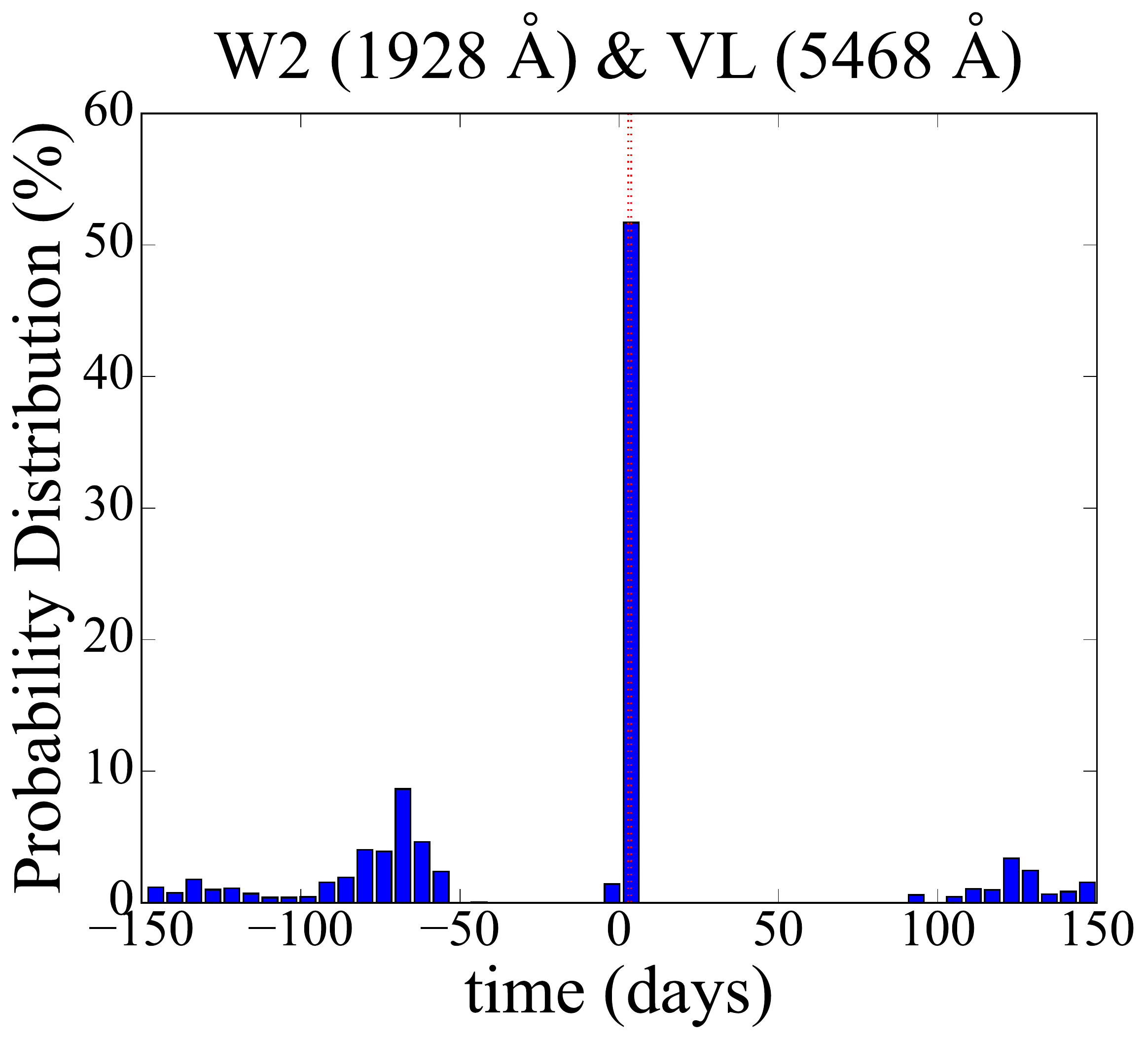} \\
            \includegraphics[width=0.3\columnwidth]{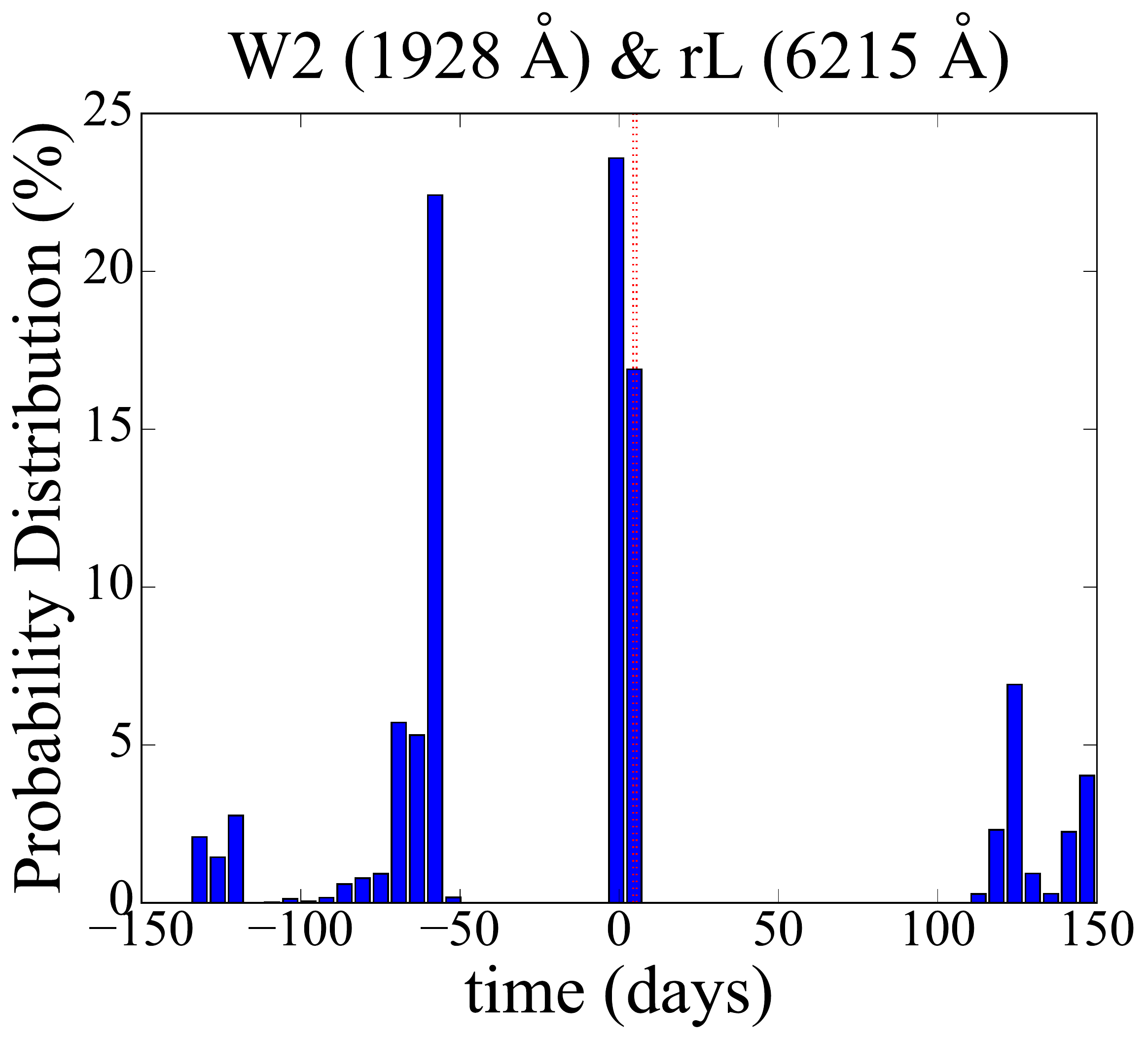} &
            \includegraphics[width=0.3\columnwidth]{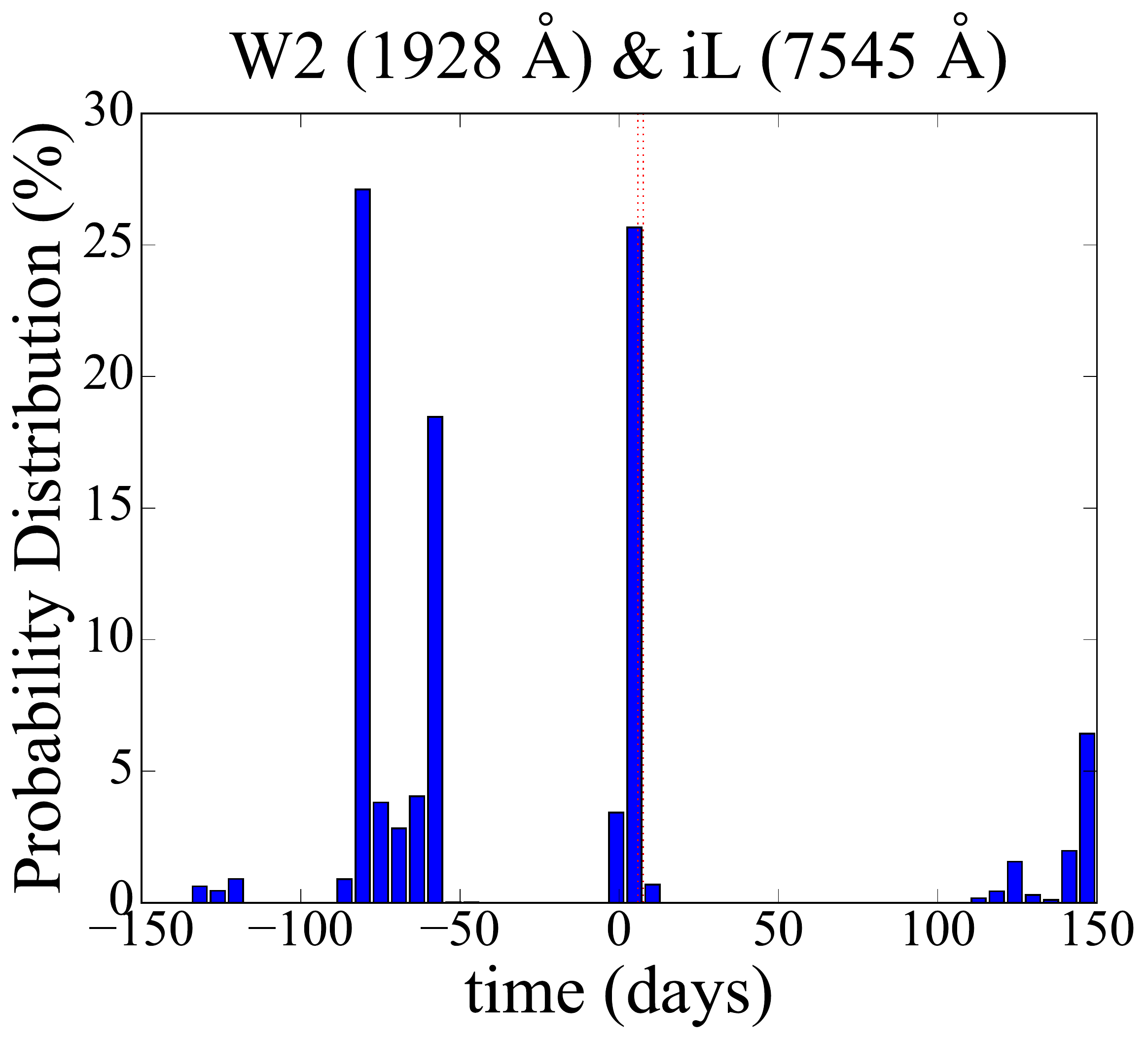} &
            \includegraphics[width=0.3\columnwidth]{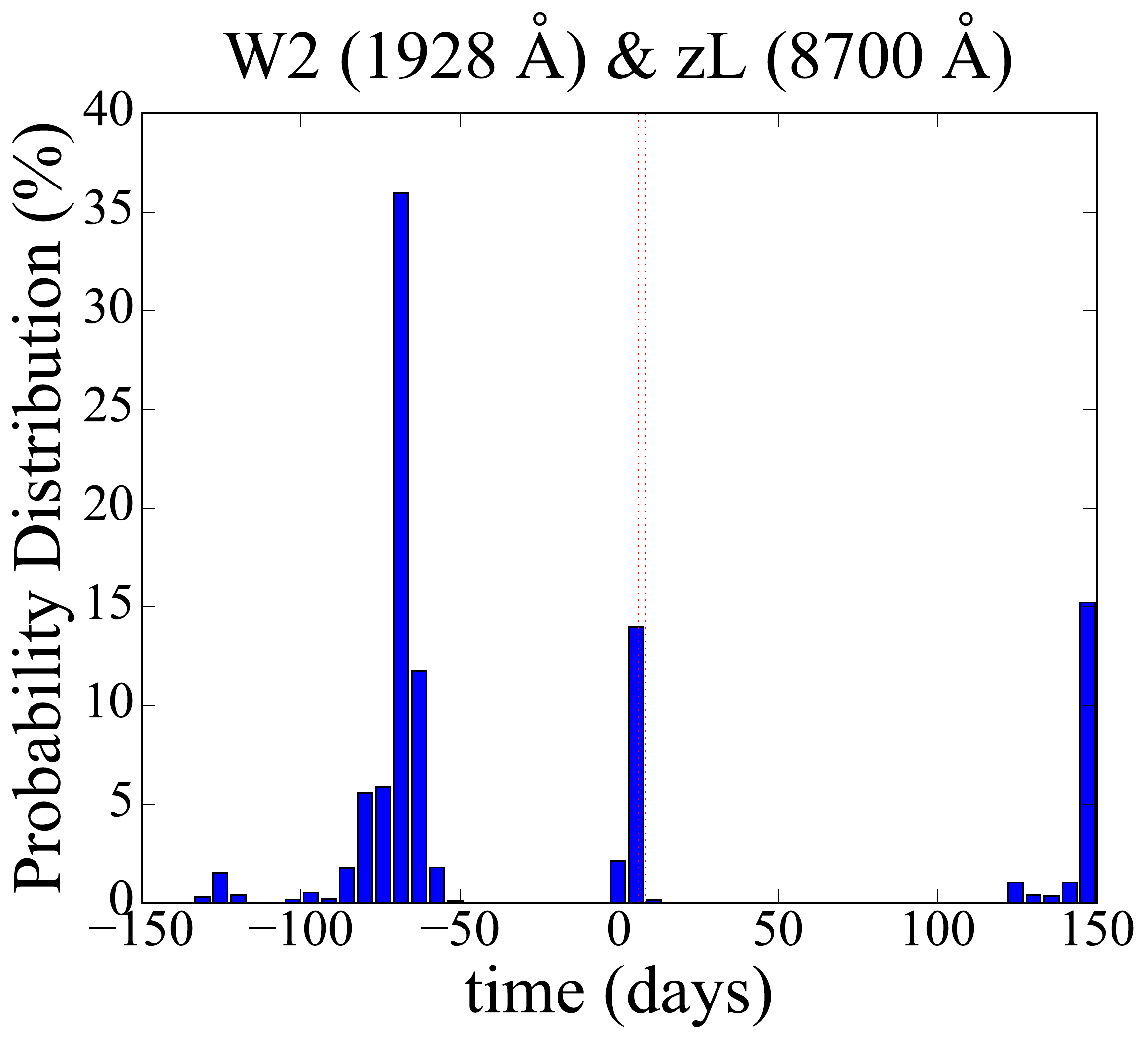} \\
            
        \end{tabular}
        \caption{Cross-correlation functions from JAVELIN showing recovered long and short lags from Fairall 9 at each band relative to the $UVW2$ band at 1982 $\rm \AA$. The red dotted lines represents the 1$\sigma$ range of the recovered short lags in \citet{F92020}. }
        \label{fig:F9_jave}
    \end{figure}
    
    \begin{figure}[H]
        \begin{tabular}{ccc}
            \includegraphics[width=0.3\columnwidth]{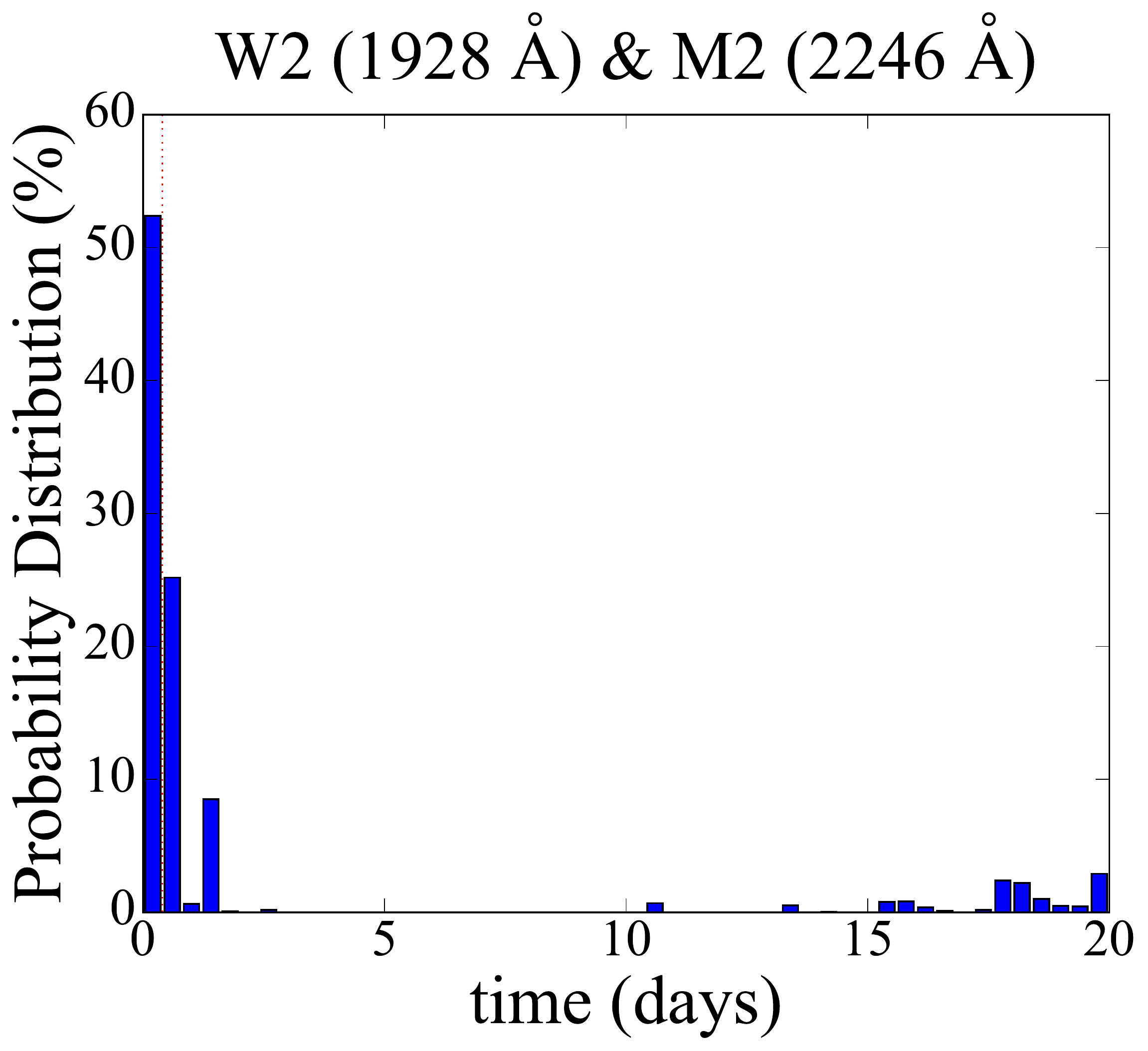} &
            \includegraphics[width=0.3\columnwidth]{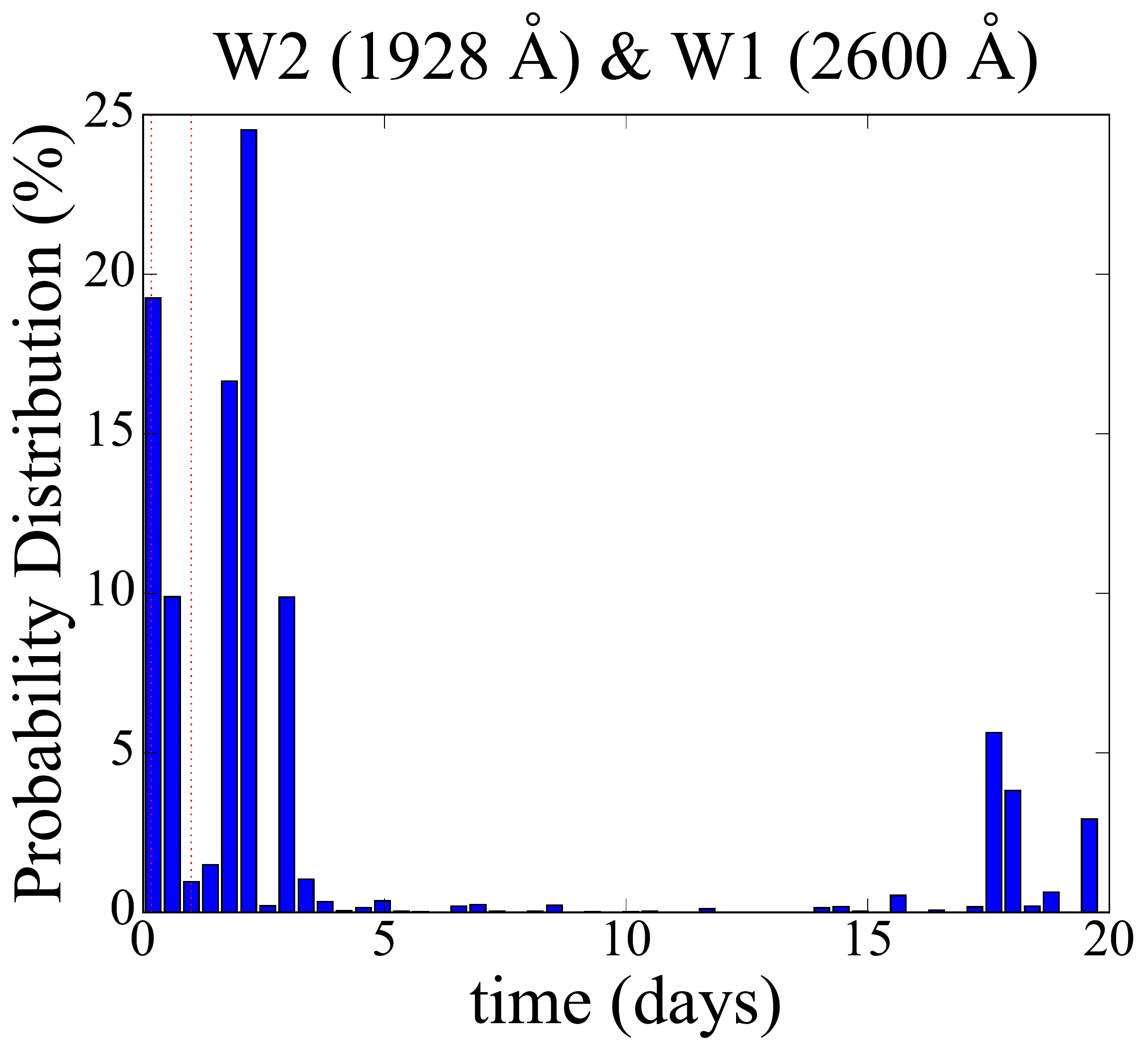} &
            \includegraphics[width=0.3\columnwidth]{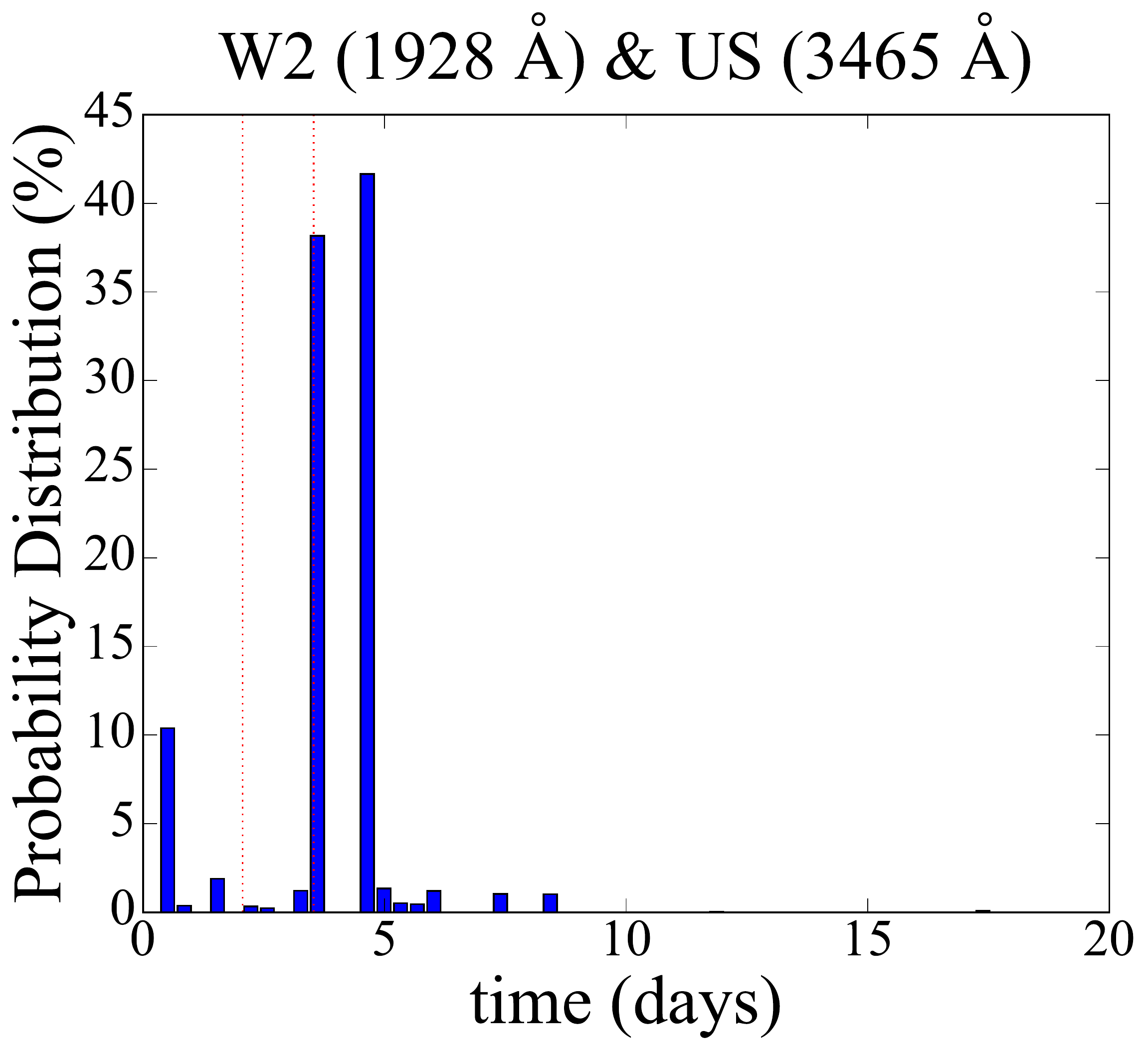} \\
            \includegraphics[width=0.3\columnwidth]{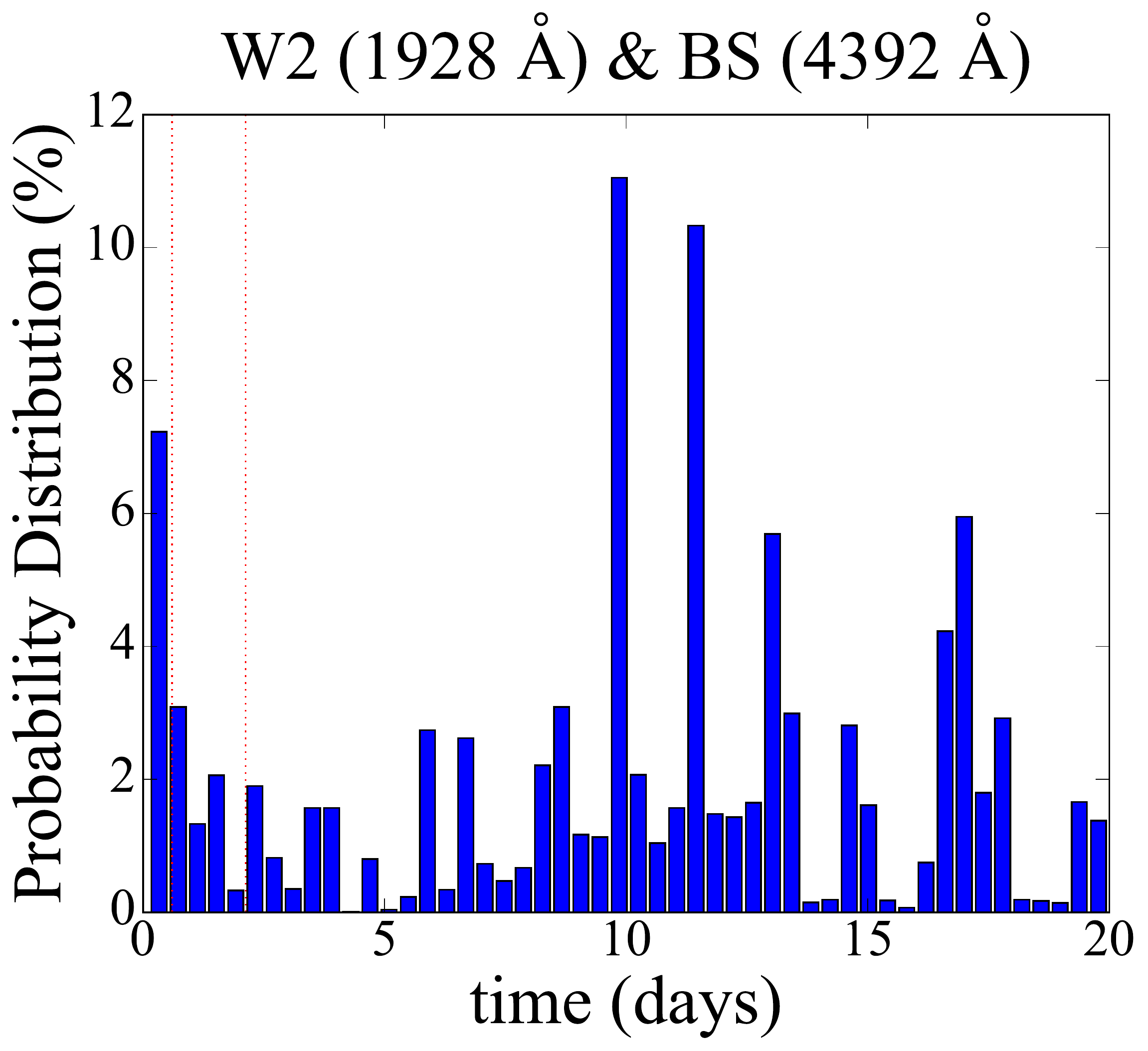} &
            \includegraphics[width=0.3\columnwidth]{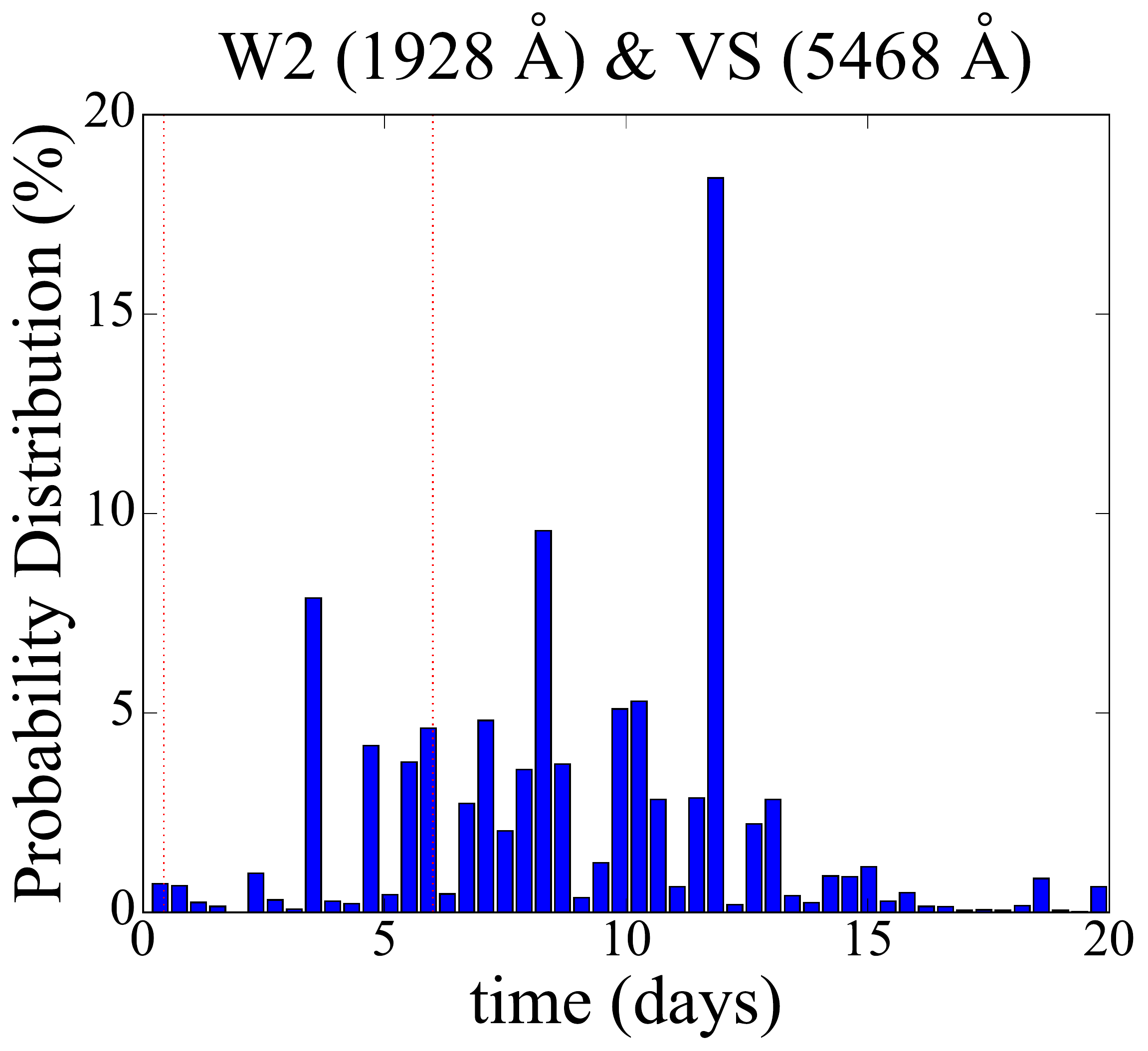} &
            \includegraphics[width=0.3\columnwidth]{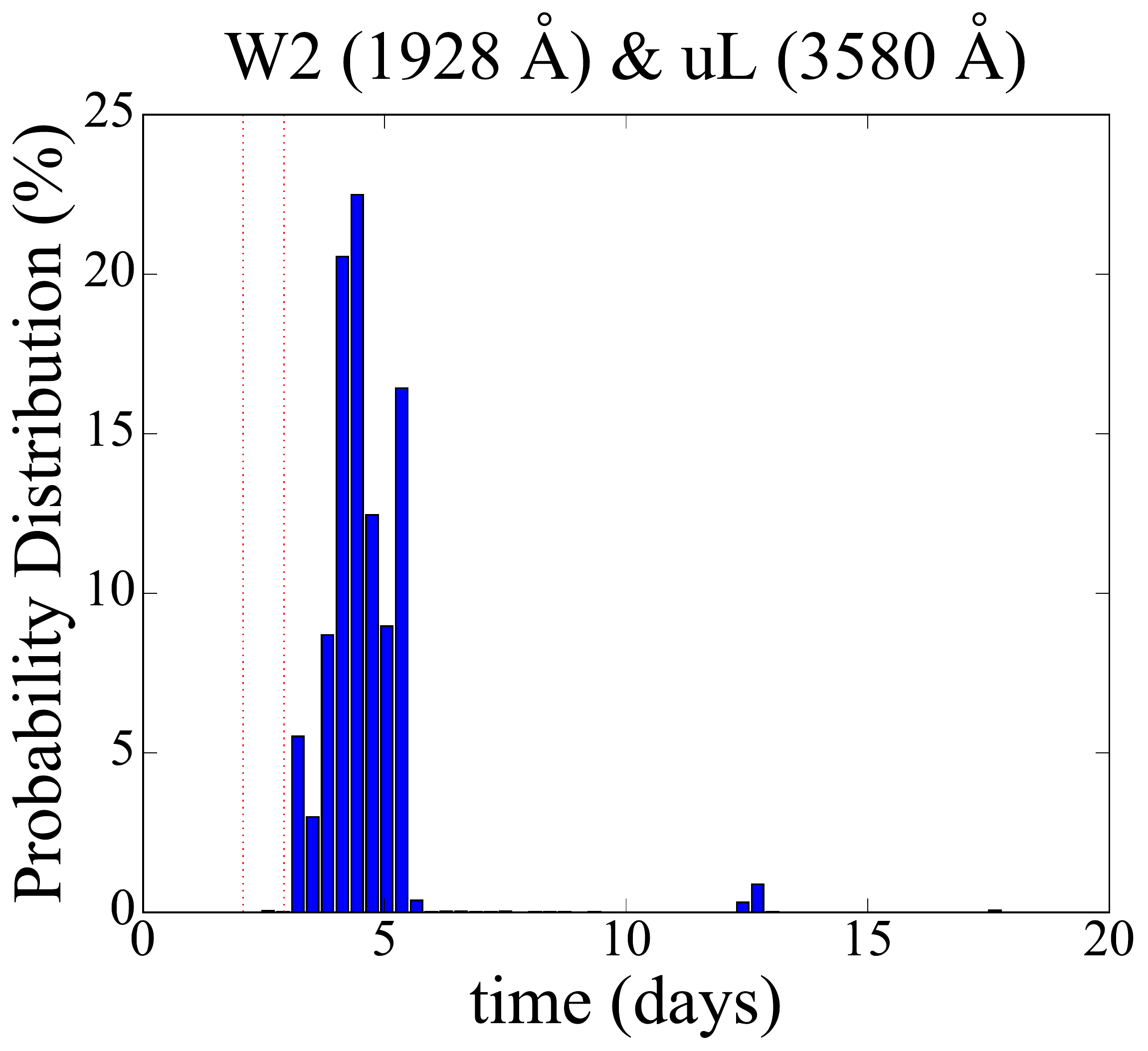} \\
            \includegraphics[width=0.3\columnwidth]{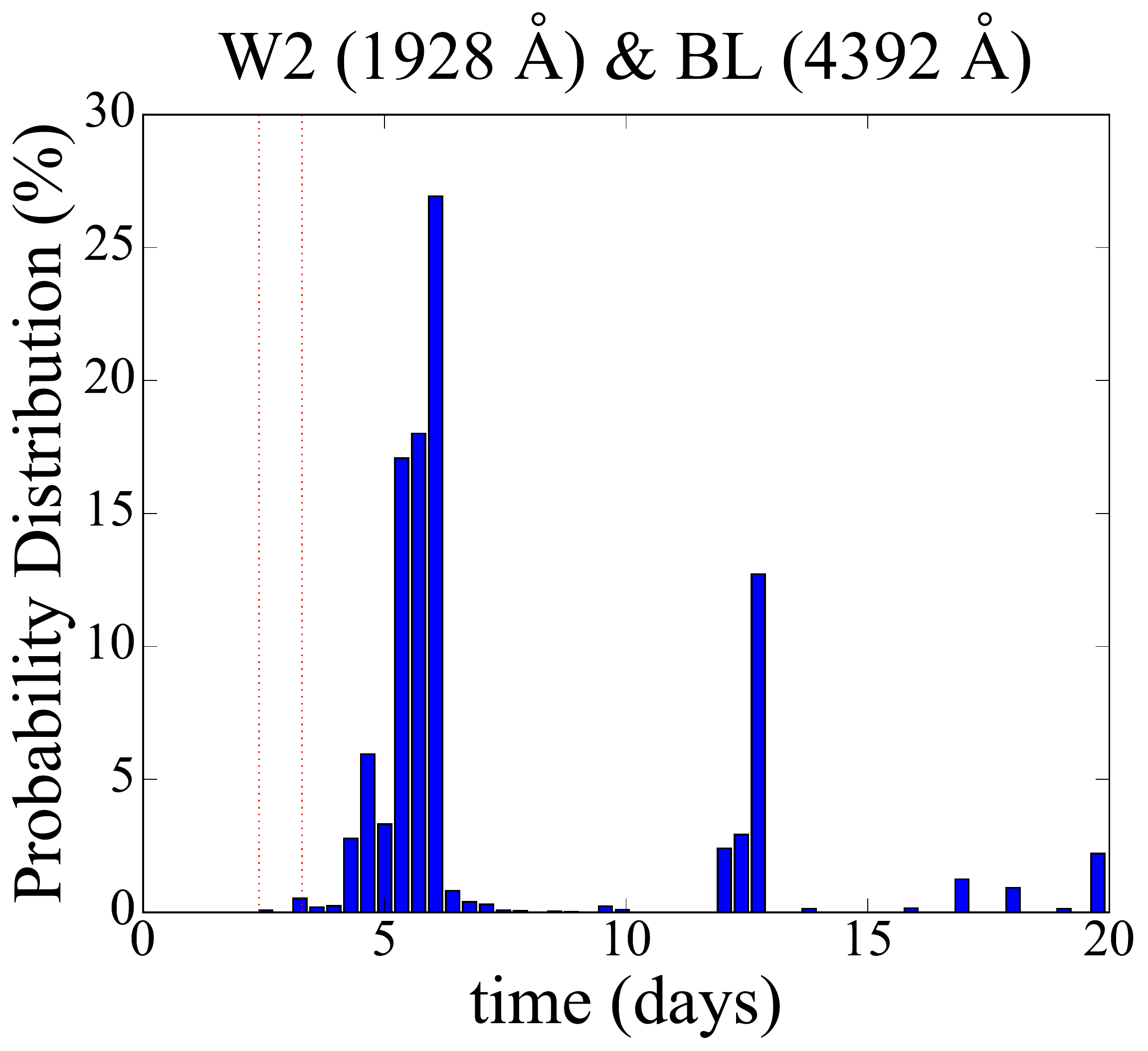} &
            \includegraphics[width=0.3\columnwidth]{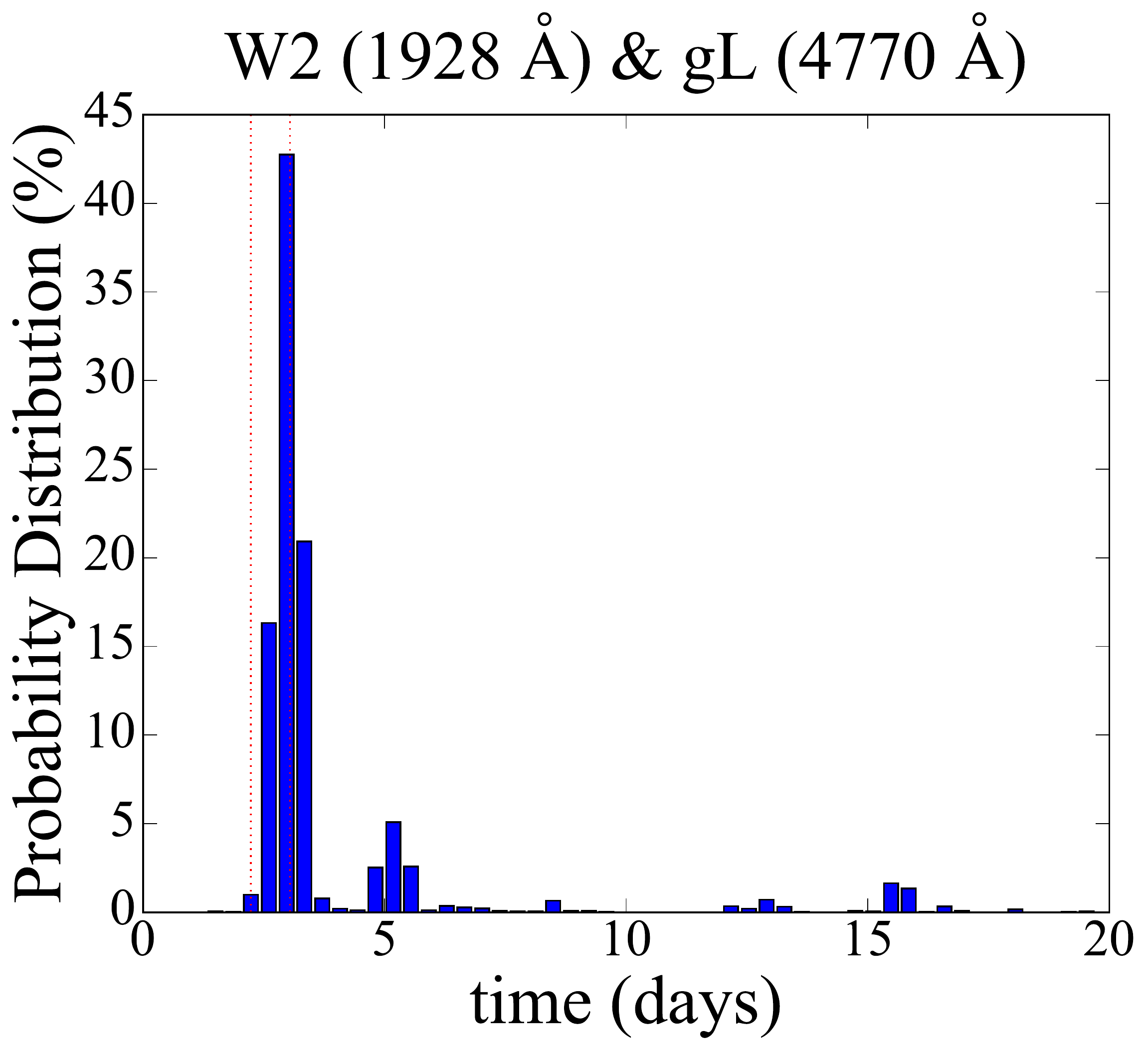} &
            \includegraphics[width=0.3\columnwidth]{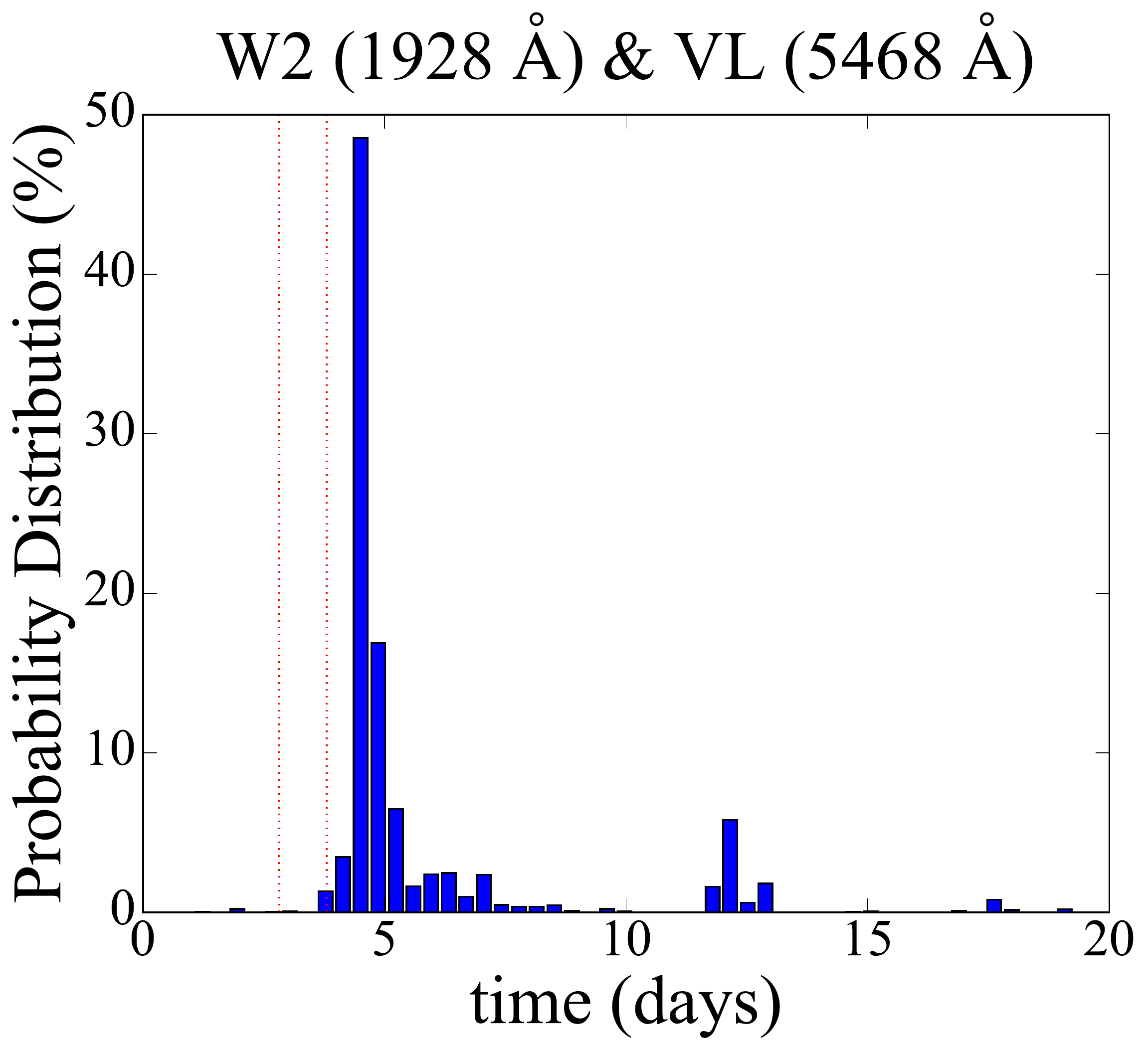} \\
            \includegraphics[width=0.3\columnwidth]{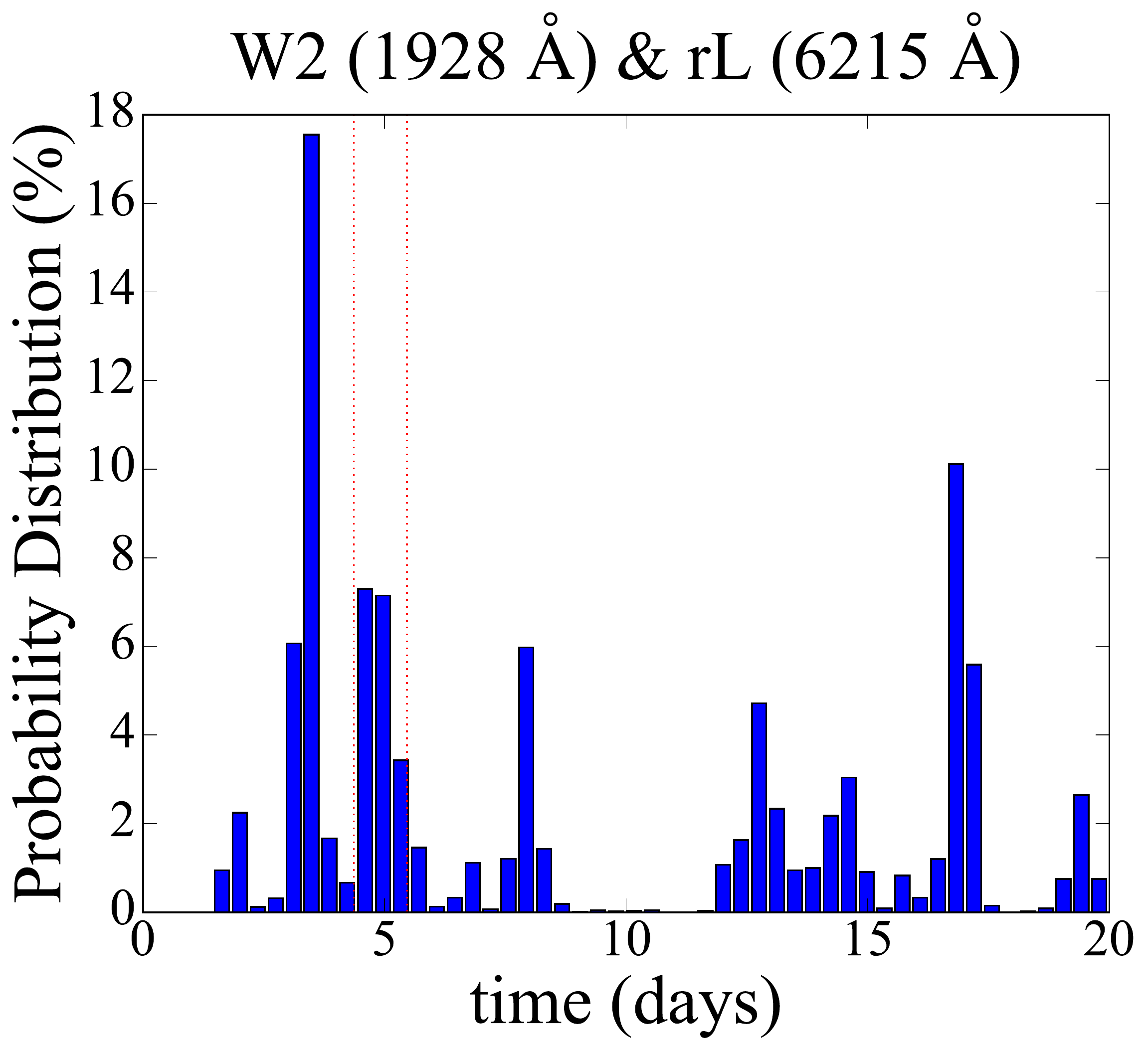} &
            \includegraphics[width=0.3\columnwidth]{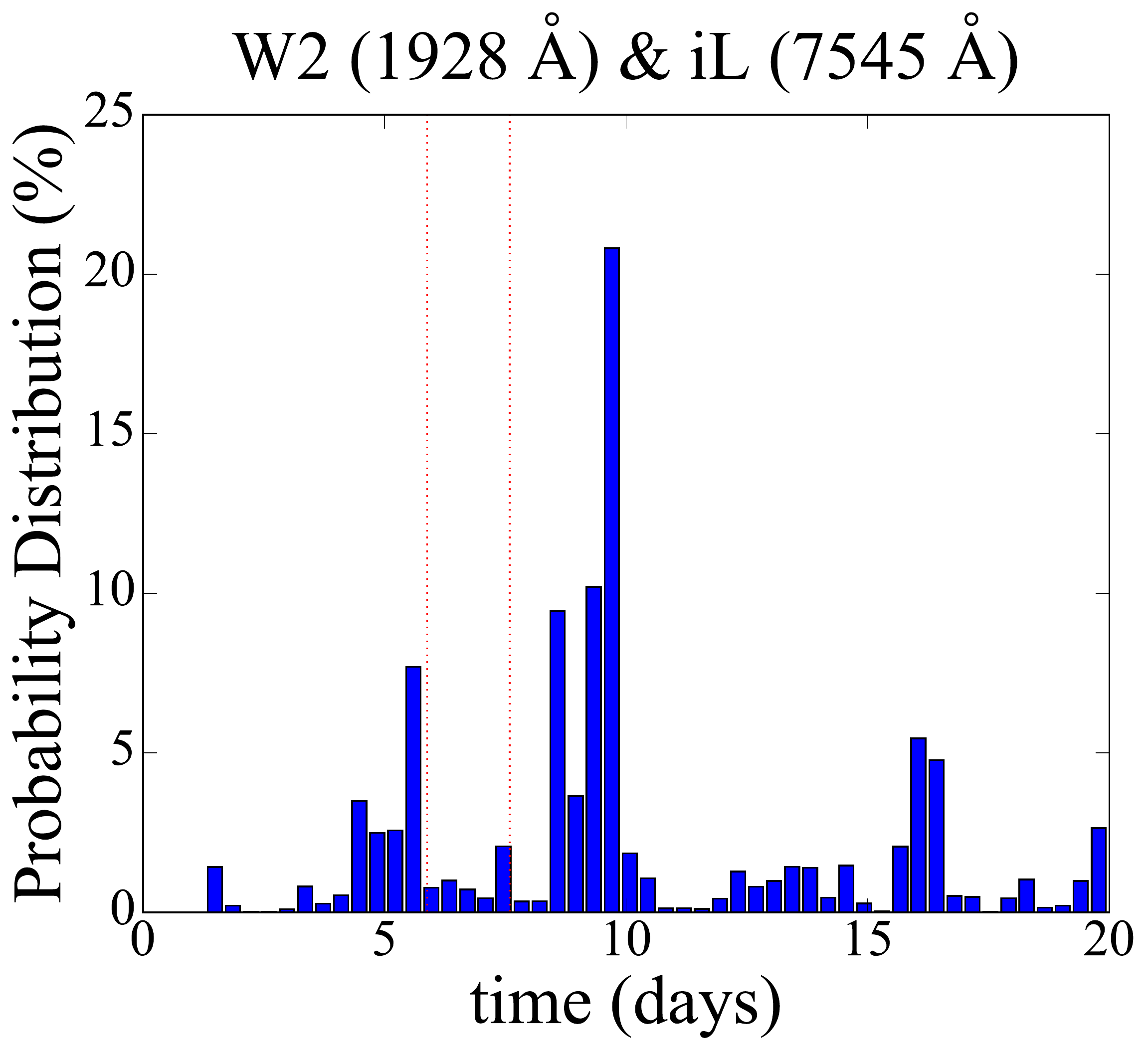} &
            \includegraphics[width=0.3\columnwidth]{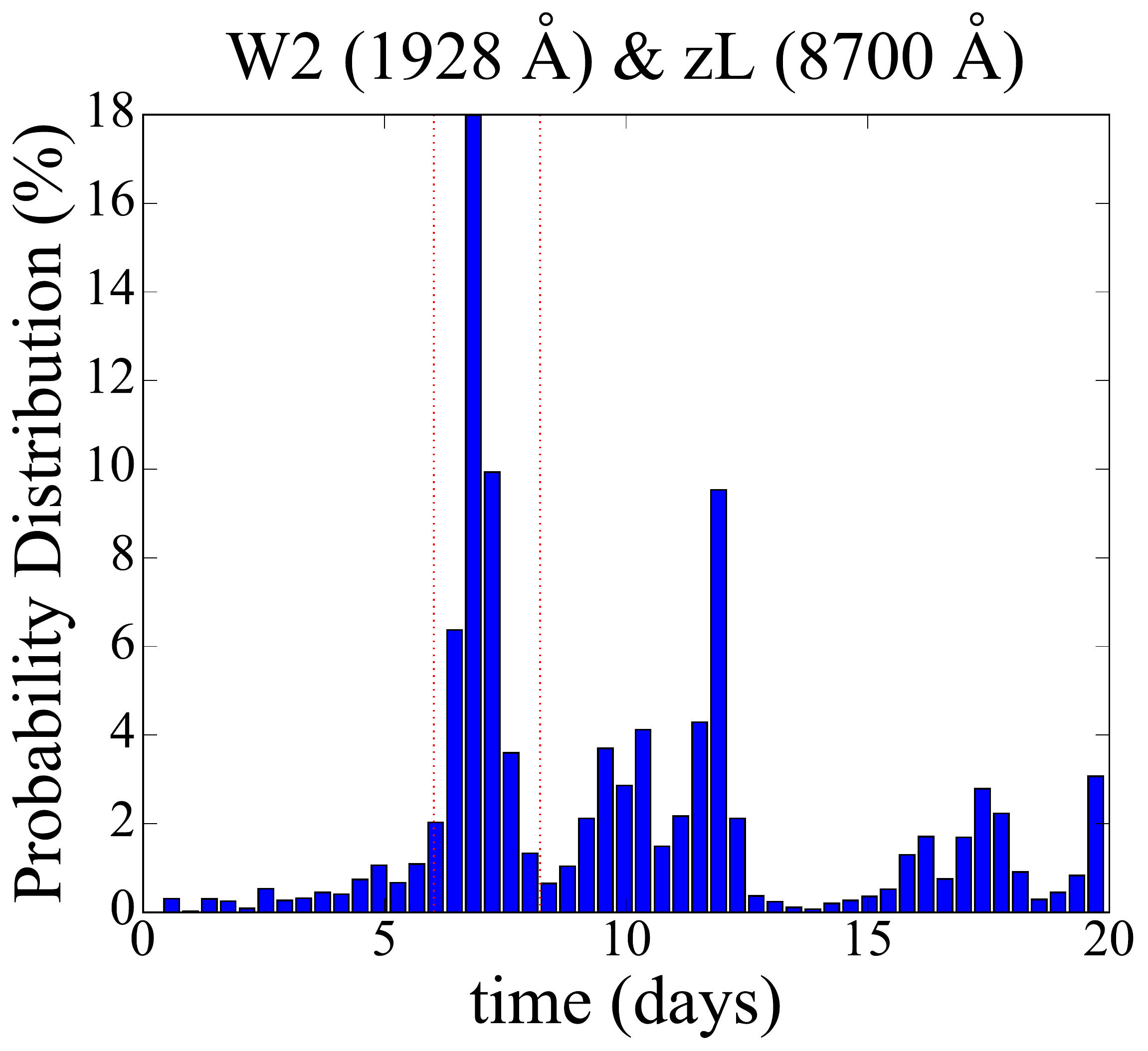} \\
            
        \end{tabular}
        \caption{Cross-correlation functions from detrended JAVELIN light curves showing recovered short lags from Fairall 9 at each band relative to the $UVW2$ band at 1982 $\rm \AA$. The red dotted lines represents the 1$\sigma$ range of the recovered short lags in \citet{F92020}.}
        \label{fig:F9_de_jave}
    \end{figure}
    
    \begin{figure*}[ht]
        \centering
        \includegraphics[width=\textwidth]{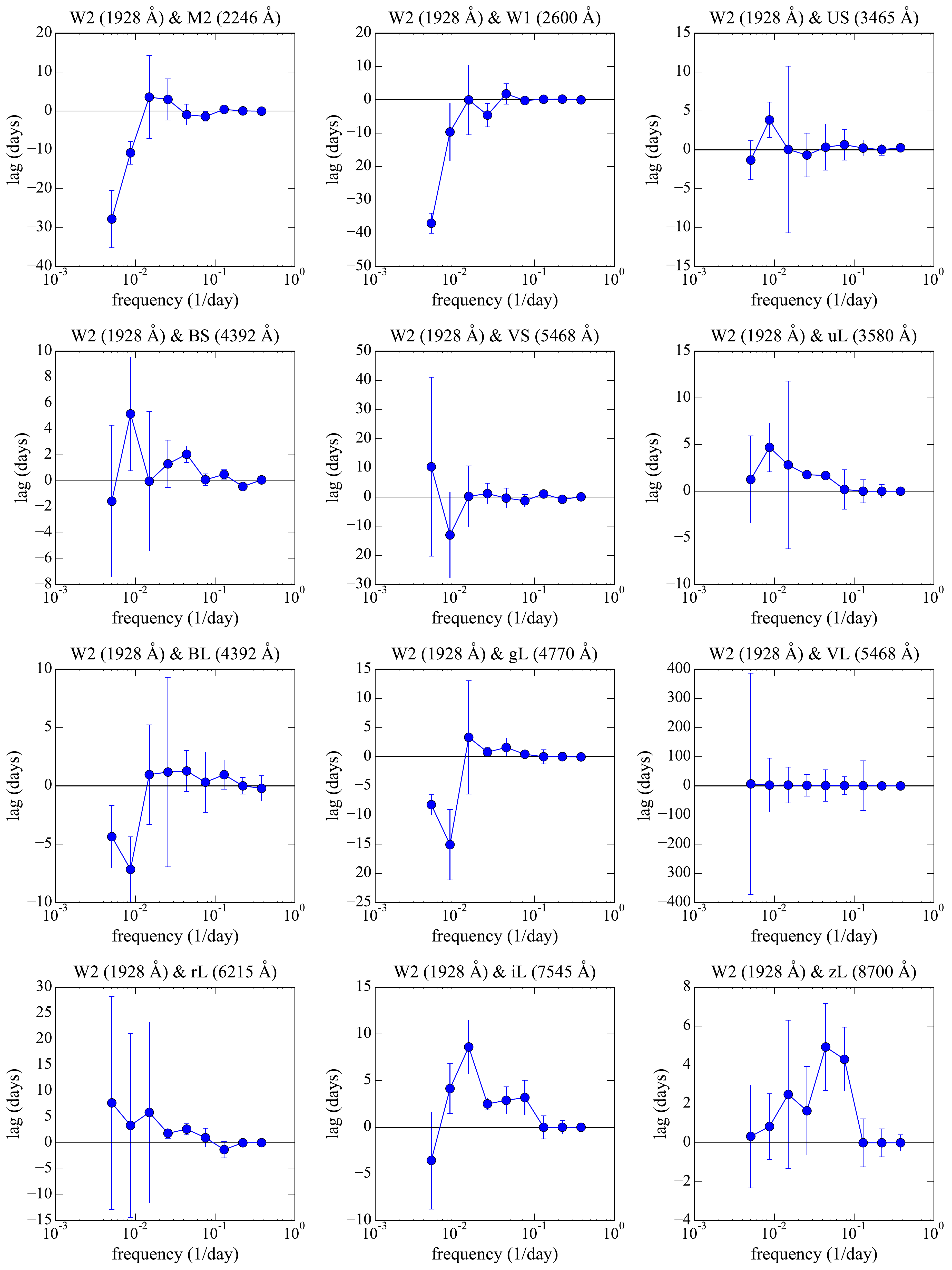}
        \caption{Lag distributions in frequency space with original light curve from \citet{F92020}.}
        \label{fig:Fourier_lags}
    \end{figure*}
    
    \begin{figure*}[ht]
        \centering
        \includegraphics[width=0.5\textwidth]{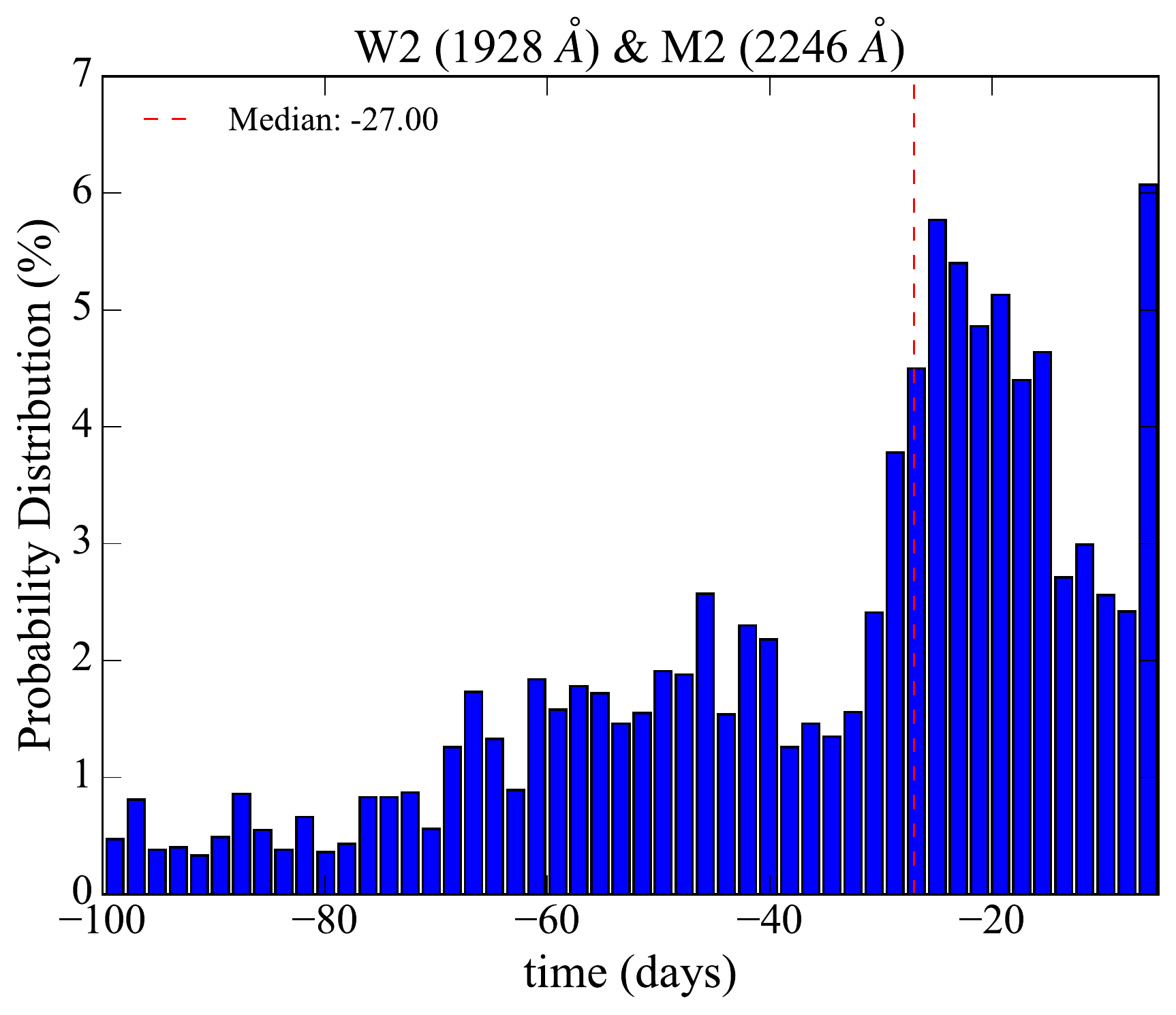}
        \caption{Cross-correlation functions from  JAVELIN between the reference $UVW2$ band at 1982 $\rm \AA$ and the the $M2$ band at 2246 $\rm \AA$. Here, we limit JAVELIN to only search for negative lags, because the power of the short lag significantly dominates over the long lag at this wavelength.}
        \label{fig:M2_neg}
    \end{figure*}

    \begin{figure*}[ht]
        \centering
        \includegraphics[width=0.7\textwidth]{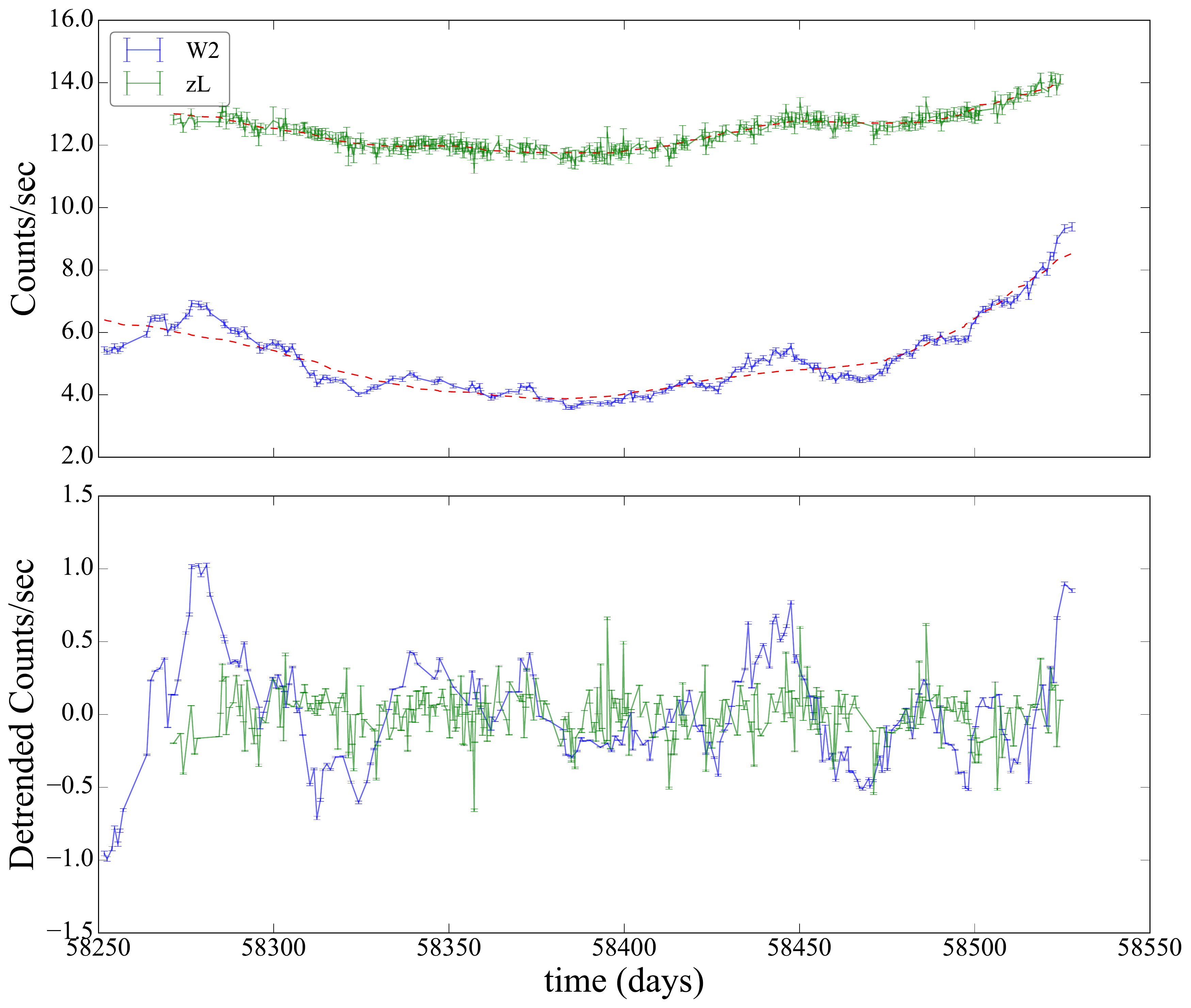}
        \caption{Top panel: example real light curves from \citet{F92020} for Fairall 9 in the reference $UVW2$ band (blue) and the longest wavelength band $z_s$ (green). Parabolas for detrending are plotted in red. Bottom panel: light curves after detrending to obtain JAVELIN short lags.}
        \label{fig:lcs}
    \end{figure*}

\end{CJK*}
\end{document}